\documentclass [11pt]{article}
\usepackage{amsmath,amsthm,amsfonts,amscd,eucal,latexsym,amssymb} 
\usepackage{epsfig}
\def\spa{\hskip -3pt} 
\def\bF{{\bf F}} 
\def\bP{{\bf P}} 
\def\sF{\mathsf{F}} 
\def\sP{\mathsf{P}}

\newsymbol\bt 1202           
\def\bC{{\mathbb C}}           

\def\bN{{\mathbb N}} 
 
\def\bR{{\mathbb R}} 
\def\bS{{\mathbb S}}

\def\bZ{{\mathbb Z}} 
 
\newsymbol\rest 1316         
 
\def\gD{{\mathfrak D}} 
 
\def\gF{{\mathfrak F}}

\def\beq{\begin{eqnarray}}
\def\eeq{\end{eqnarray}}
\def\pa{\partial}
\def\at{\left(}               
\def\aq{\left[}               
\def\ag{\left\{}              
\def\ct{\right)}              
\def\cq{\right]}              
\def\cg{\right\}}             
\newcommand{\ca}[1]{{\cal #1}}         

\def\ka{\kappa}

\begin{document} 
 
\hfill{\sl Preprint  UTM 644 - July 2003} 
\par 
\bigskip 
\par 
\rm

 
\par 
\bigskip 
\LARGE 
\noindent 
{\bf Quantum Virasoro algebra with central charge $c=1$ on the  horizon of a $2D$-Rindler spacetime.} 
\bigskip 
\par 
\rm 
\normalsize 
 
 
\large 
\noindent {\bf Valter Moretti$^{1,2,a}$} and {\bf Nicola Pinamonti$^{b}$} \\
Department of Mathematics, Faculty of Science,
University of Trento, \\ via Sommarive 14, 
I-38050 Povo (TN), 
Italy\\\par 
\smallskip\noindent 
$^1$  I.N.d.A.M.,  Istituto Nazionale di Alta Matematica ``F.Severi'',  unit\`a locale  di Trento\\ 
$^2$  I.N.F.N., Istituto Nazionale di Fisica Nucleare,  Gruppo Collegato di Trento\\
$^a$ E-mail: moretti@science.unitn.it,  $^b$ E-mail: pinamont@science.unitn.it\\ 

\rm\large\large

\rm\normalsize 

\rm\normalsize

 
\par 
\bigskip 
 
\noindent
\small 
{\bf Abstract}.  
Using the holographic machinery built up  in a previous work, we show that 
the hidden $SL(2,\bR)$ symmetry of a scalar quantum field propagating in 
a Rindler spacetime admits an enlargement in terms of a 
unitary positive-energy  representation of
Virasoro algebra defined in the Fock representation. That  representation has central charge $c=1$. 
The Virasoro algebra of operators gets a manifest geometrical meaning 
if referring to the holographically associated QFT on the horizon:
It is nothing but a representation of the algebra of vector fields defined on the horizon 
equipped with a point at infinity.  All that happens provided the Virasoro ground energy $h:=\mu^2/2$ vanishes and,  in that case,  the
 Rindler Hamiltonian is associated with a certain Virasoro
generator.  If a suitable regularization procedure is employed,   for  $h=1/2$,  the ground state of that generator 
corresponds to thermal states when examined in the Rindler wedge, taking the expectation value 
with respect to Rindler time. This state has inverse temperature
$1/(2\beta)$, where  $\beta$ is the parameter used to define the initial $SL(2,\bR)$ unitary 
representation. (As a consequence the restriction of Minkowski vacuum to Rindler wedge is obtained
by fixing $h=1/2$ and  $2\beta=\beta_U$, the latter being Unruh's inverse temperature).
Finally, under Wick rotation in Rindler time, the pair of QF theories which are built up on the future and past horizon 
defines a proper two-dimensional conformal quantum field theory on a cylinder. 
\normalsize
\newpage

\section{Introduction and summary of previously obtained results.}

\noindent {\bf 1.1}. {\em Introduction}. 
A number of papers has been concerned with the issue of the statistical origin of black-hole entropy. 
Holographic principle \cite{thoo93, thoo95, suss95} arose by the idea  
that gravity near the horizon should be described by a low dimensional theory with a higher dimensional group
of symmetry.  Maldacena and Witten \cite{mald98,witt98} 
showed that there is a correspondence between quantum field theory in
an asymptotically $AdS$ spacetime,  the ``bulk'',
 and a conformal theory on its ``boundary'' at spacelike infinity.  
Rehren proved rigorously some holographic results for free quantum fields in a $AdS$ 
background, establishing a correspondence between bulk observables and boundary observables 
without employing string machinery \cite{rehr00a,rehr00b}. 
Dealing with QFT in $2D$-Rindler spacetime, we have proved in a recent work \cite{mopi03}, that it is possible to 
define a free quantum theory on the horizon of a two dimensional Rindler space. 
That theory enjoys holographic interplay with the analogous theory defined in the bulk.
More precisely, there are two holographic theorems. 
The former shows that there is a $*$-algebra injective homomorphism from the 
algebra of the bulk observables associated with the Rindler free field to  the 
algebra of the horizon observables associated with the horizon free field.
The latter identifies the observables of the theories
form the point of view of unitary equivalences whenever the theory is 
represented in suitable Fock spaces (in that case also the vacuum states are in correspondence
thorough the unitary operator which realizes holography).
An interesting consequences is that the
``hidden'' $SL(2,\bR)$  symmetry of 
free quantum field theory  in the bulk found in \cite{mopi02} becomes manifest when transposed on the Killing horizon
by means of unitary holography. In fact, due to the spectrum of the Hamiltonian operator,  QFT theory 
in the bulk turns out
to be invariant under a unitary representation of $SL(2,\bR)$ but such a quantum symmetry cannot be induced
by the geometric background because the isometries of Rindler space has a Lie algebra
different form that of $SL(2,\bR)$ \cite{mopi02}. Nevertheless, the unitary representation of $SL(2,\bR)$
which realizes that bulk symmetry becomes manifests, i.e. it reveals a clear geometric meaning,
 if it is examined on the horizon by means of  the holographic machinery.  All that is summarized within subsection 1.2  in some details.\\
Overlap with ideas and results of \cite{mopi03} 
is present in the literature especially due to  Schroer \cite{SC},  Schroer and Wiesbrock \cite{SW00},
 Schroer and Fassarella \cite{SF01}. In those papers an approach to holography 
similar to ours is implemented in the framework of LightFront Holography
developed at algebraic level using nets of local observable algebras. 
  From a very elementary point of view, a relevant difference with our machinery
 is the fact that the quantization of the bulk field used by Schroer and collaborators 
 is that referred to Minkowski vacuum and Minkowski time
 instead of Rindler vacuum and time.  From a pure physical point of view, perhaps, the quantization
with respect Rindler frame is more interesting if one tries to use our machinery as a starting point to investigate
QFT near the bifurcate Killing horizon of a black hole: Rindler quantization corresponds to quantization 
in a reference frame that gives rise to Minkowski coordinates far from the black hole and the associated particles 
should be those things are made of.  
  However the interplay of Schroer and collaborators' ideas and achievements 
and  procedures and results  presented in our paper deserves further investigation.
Another relevant paper which merits particular  quotation  is that  by  Guido, Longo, Roberts and Verch \cite{GLRV01}. Overlap with some results arising by our approach are present in section
4 of \cite{GLRV01}. In that section, in the very general context of QFT in curved spacetime in terms of nets of local $C^*$  algebras  (and Von Neumann representations) and making use of very general theorems by Wiesbrock  on local quantum field theory defined on $\bS^1$ and covariant with respect to $PSL(2,\bR) :=SL(2,\bR)/\pm I$,  it is proven the existence of a local quantum field theory (covariant with respect to $PSL(2,\bR)/\pm I$)  defined on the bifurcate Killing horizon. This is done by considering a net of Von Neuman algebras in the representation of a state which is, in restriction to the subnet of observables which are localized at the horizon, a KMS state at Hawking temperature for the Killing flow.\\
In \cite{mopi03} we found some clues for the existence of a whole unitary representation of Virasoro algebra which extends the $SL(2,\bR)$ unitary representation on the horizon.
In this paper we prove the very existence of a full unitary representation of Virasoro algebra 
with central charge $c=1$ for quantum field theory defined on the horizon.
That fact is interesting for several reasons in relation with the problem of the 
statistical interpretation of black hole entropy. In fact, there are several attempts to 
give a statistical explanation to black hole entropy by counting microstates
in terms of the degeneracy  of an eigenspace of a certain Virasoro generator in a 
suitable  irreducible unitary representations of
Virasoro algebra \cite{stro-etall, carlip-etall}. This is done  by means of the so-called ``Cardy's formula''. 
These approaches are, in fact, based on the existence of a Virasoro algebra (with central charge $c\neq 0$)  
 in terms of generators of diffeomorphisms of the black hole manifolds considering the 
horizon as a boundary.
The algebra  of  the associated  generators in the 
Hamiltonian ADM formulation of gravity gets a non-vanishing central charge. 
Under the supposition that  a quantum version of that Virasoro representation exists, that the value of central charge is not affected by the quantization procedure
and that the actual value of the black hole mass is a eigenvalue of the Virasoro generator $L_0$,
it is possible to compute  the degeneracy of that eigenspace  by means of Cardy's formula
because of the presence of a central charge.
The logarithm of the degeneracy gives the very black-hole entropy law barring logarithmic corrections.\\
The main problem of all of those approaches is that the Virasoro algebra representation with 
non vanishing central charge is proven to exists at classical level only in the Hamiltonian formulation.
All  derivations of black hole entropy by that way are based on the found classical formulas
and on the supposition that there is a quantum version of the found Hamiltonian structure (in order to  use Cardy's formula). \\
To make contact with the content of this paper where 
a quantum scalar field propagating in a $2D$-Rindler space is considered, we notice that
 in the approaches outlined above,  the only near-horizon structure 
is sufficient to use the Virasoro-Cardy machinery  \cite{carlip-etall,stro-etall,giacomini-pinamonti}.  Moreover, for a Schwarzschid black-hole manifold,
the relevant algebra of   diffeomorphisms  is that of  diffeomorphisms in the plane $r,t$ 
 which preserve the horizon  structure.  Hence it seems  that
 2D-Rindler models are relevant to this context.
 On the other hand, a scalar field arises naturally in these 2D-Rindler space approaches 
by dimensional reduction \cite{giacomini-pinamonti} form the gravitational theory in $4D$ in the presence of
spherical symmetry. That field supports information of part of $4D$-dimensional gravity 
in the $2D$ model. 
Concerning the problem of the existence of a Virasoro representation at quantum level we stress that, in this paper,
 we prove  that a very positive-energy unitary representation of Virasoro algebra does exist at quantum level
for the quantum field defined on the horizon.  That algebra of operators can be defined also for the scalar field propagating in the bulk via unitary holography.\\
In fact,  Section 2 of this  paper is devoted to show that 
the bulk hidden $SL(2,\bR)$ symmetry admits an enlargement in terms of a 
unitary  representation of
Virasoro algebra with central charge $c=1$ defined in  Fock representation. 
The Virasoro algebra of operators gets a manifest geometrical meaning 
if referring to the holographically associated QFT on the horizon:
It gives rise to a unitary representation of a group of automorphisms  
  of the  $*$-algebra  generated by  field operators. This representation 
  is  induced by group of the diffeomorphisms of the horizon compactified by adding
  a point at infinity. 
Moreover, a sub-representation which is generated by three certain Virasoro generators reduces to the $SL(2,\bR)$
representation previously found. Under Wick rotation with respect to Rindler time, the pair of QF theories
which are built up on the future and past horizon defines a proper two-dimensional 
conformal quantum field theory. That CFT can be realized  on the Riemann surface given 
by a two-dimensional cylinder.  In Section 3 we prove that, with a suitable choice of the weight of the found Virasoro algebra of operators,
a certain generator which generalizes Rindler Hamiltonian, admits a ground state $\Psi$ which
enjoys notable thermodynamic properties: When that state is examined  in the bulk via holography,
and a suitable mean is computed with respect to Rindler-time evolution,  $\Psi$ reveals itself  as a thermal 
state whose inverse temperature is $2\beta$. $\beta$ being the parameter initially used 
to build up the unitary $SL(2,\bR)$ representation in the bulk.\\

 \noindent {\bf 1.2}. {\em QFT on a Killing  horizon, unitary and algebraic holographic theorems and $SL(2,\bR)$
  symmetry}. We summarize part of the content of \cite{mopi03} relevant for this work within the following five steps.
  
{\bf [a]} Consider the globally-hyperbolic spacetime ${\bf R}$ called  two-dimensional {\em Rindler wedge} with metric
$ds^2_{\bf R} = -{\kappa}^2 y^2 dt^2 + dy^2$, 
 that can be obtained by a suitable near horizon approximation of a 
 general Schwarzschild-like metric also dropping the angular coordinates \cite{mopi03},
 above $t\in \bR$, $y\in (0,+\infty)$ are global coordinates.
A free Klein-Gordon scalar field $\phi$  in ${\bf R}$ satisfies the equation of motion
$-\partial^2_t \phi  + {\kappa}^2 \left(y \partial_yy \partial_y  -  
 y^2 m^2 \right) \phi =0$. In Rindler quantization, the one-particle Hilbert space ${\cal H}$  
 consists of the space of complex linear combinations of the positive frequency parts
 of smooth real solutions $\psi$ of the K-G equation 
 with compact Cauchy data. 
The natural {\em symplectic form}  on that space is
$ \Omega({\psi},\psi') := \int_{\Lambda} \left(\psi' \nabla^\mu {\psi}
    - {\psi} \nabla^\mu \psi'\right)  
n_\mu\: d\sigma$, 
$\Lambda$ being any Cauchy surface,  $d\sigma$ the induced measure and   $n$ a unit future-oriented 
normal vector.
 Every $\psi$ decomposes into $\partial_t$-stationary modes as
\begin{eqnarray}
\psi(t,y) = \psi_+(t,y) + c.c. =
\int_{0}^{+\infty} \: e^{-iEt}\sum_\alpha\Phi^{(\alpha)}_{E}(y) \tilde\psi^{(\alpha)}_+(E)\: dE + c.c. \label{wave}
\end{eqnarray}
The index $\alpha$ distinguishes between two cases: if $m>0$ there is a single mode 
$\Phi^{(\alpha)}_{E}=\Phi_{E}$.
If $m=0$ there are two values of $\alpha$, corresponding to 
$in$going  and $out$going modes, $\Phi^{(in)/(out)}_E= e^{\mp iE\ln{(\ka
y)/\ka}}/\sqrt{4\pi E}$. In that case both ingoing and outgoing components in (\ref{wave}) must have Cauchy compact support.
The one-particle Hilbert space of wavefunctions is obtained by taking the completion of the space of complex linear combinations of
positive frequency wavefunctions (obtained from Cauchy support compactly real wavefunctions) with respect to the 
Hermitean scalar product $-i\Omega(\overline{\psi'_{+}},\psi_{+})$.
$\partial_t$ evolution of a wavefunction $\psi$ is equivalent to the action of the one-parameter subgroup
generated by an Hamiltonian $H$ on the associated $\psi_+\in {\cal H}$. $\sigma(H)=[0,+\infty)$ for $m\geq 0$.
If $m>0$ there is no energy degeneration and the one-particle Hilbert space $\ca{H}$ is isomorphic to $L^2(\bR^+,dE)$
via spectral decomposition of $H$.
In the other case ($m=0$), twofold degeneracy implies that $\ca{H} \cong L^2(\bR^+,dE)\oplus L^2(\bR^+,dE)$. 
Let us pass to the bosonic Fock space, ${\gF}({\cal H})$, associated with ${\cal H}$.
The  {\em quantum field $\Omega(\:\cdot,\hat \phi)$} of our theory  is the map
\begin{eqnarray}
 \psi \mapsto \Omega(\psi, \hat\phi) &:=& ia(\overline{{\psi}_+}) -ia^{\dagger}(\psi_+) \label{PHI}\:, 
\end{eqnarray}
 where $\psi$ is any real compactly supported wavefunction  and $a(\overline{\psi_+})$ and $a^{\dagger}(\psi_+)$
 respectively denote the annihilation  and construction operator associated with the one-particle states $\overline{\psi_+}$
and $\psi_+$ respectively,  defined in ${\gF}({\cal H})$ and  referred to the Rindler vacuum 
  $|0\rangle$ (that is $|0\rangle_{in}\otimes |0\rangle_{out}$ 
if $m=0$).
$\Omega(\psi, \hat\phi)$ is essentially self-adjoint in the dense invariant subspace 
spanned by all states containing a finite  arbitrarily large number of particles with states 
given by positive-frequency  wavefunctions. 
 Every wavefunction $\psi$ in (\ref{wave})
can be obtained as $\psi = E(f)$ where $f$ is an associated compactly supported smooth function in ${\bf R}$
and  $E$ is the {\em causal propagator} (the ``advanced-minus-retarded''
two point function) of Klein-Gordon operator. Moreover 
\begin{equation}\int_{\bf R} \psi f \:d\mu_g = \Omega(Ef,\psi) \:\:\mbox{and}\:\:
\int_{\bf R} h(x)(Ef)(x) \: d\mu_g(x) = \Omega(Ef,Eh)
\label{psif}\:, 
\end{equation}
 $\mu_g$ being the measure induced by the metric in ${\bf R}$.
(\ref{psif}) suggests to define \cite{Wald} a  quantum-field operator 
 smeared  with  compactly-supported complex-valued  functions $f$, as the linear map
\begin{eqnarray}
f\mapsto \hat\phi(f) := \Omega(Ef, \hat\phi)\:,\label{fsmeared}
\end{eqnarray}
which  is formally equivalent 
to the non-rigorous but popular definition
\beq
\hat{\phi}(t,y)=\int_0^\infty\sum_\alpha
e^{-iEt}\Phi^{(\alpha)}_E(y)a_{E\alpha}+e^{iEt}\overline{\Phi^{(\alpha)}_E(y)}a_{E\alpha}^\dagger dE\,.
\eeq
The  rigorous version of the formal identity $[\hat\phi(x), \hat \phi(x')] = -iE(x,x')$ is
\begin{eqnarray}
[\hat\phi(f), \hat \phi(h)] = -iE(f,h) := -i\int_{\bf R} h(x)(Ef)(x) \: d\mu_g(x)\:. \label{locality}
\end{eqnarray} 

{\bf [b]} In \cite{mopi02,mopi03} we have established that, if $m>0$, $\ca{H}$ is irreducible under a
(uniquely determined) strongly-continuous  unitary 
representation  of ${SL}(2,\bR)$ whose Lie algebra is given  by the (uniquely determined) self-adjoint extension of the 
real linear combinations of 
operators $H_0, D, C$:
\begin{equation}
H_0 := E \:,\:\:\:\:\:
D := -i\left(\frac{1}{2} + E\frac{d \:}{d E}\right)\:, \:\:\:\:\:
C := -\frac{d \:}{d E} E\frac{d \:}{d E} +
\frac{(k-\frac{1}{2})^2}{E}\:.\label{ge3F}  
\end{equation} 
$k$ can arbitrarily be fixed in $\{1/2,1,3/2,\ldots\}$.
$iH_0,iC,iD$ enjoy the commutation relations of the Lie algebra of $SL(2,\bR)$
in a suitable dense and invariant domain ${\cal D}_k$ where they, and their real linear combinations,
are essentially self-adjoint
\cite{mopi03} and $\overline{H_0}=H$. 
${\cal D}_k$ is the subspace spanned by the eigenvectors of the operator
\begin{equation}
K_\beta := \frac{1}{2}\left(\beta H_0 + \frac{1}{\beta} C\right) \label{Kbeta}\:,
\end{equation}
$\beta$ being a constant with the dimensions of an inverse energy. The unitary representation does not
depend on the value of $\beta$. The spectrum of the self-adjoint operator 
$\overline{K_\beta}$ (initially defined on ${\cal D}_k$) is a pure point spectrum without degeneracy, it 
does not depend on 
$\beta$ itself and is $\sigma(K_\beta)= \{\lambda_n\:|\: \lambda_n = n\:\:,\:\: n=k,k+1,k+2,\ldots  \}$.
If $\mathsf{L}^{(\alpha)}_p$ are modified Laguerre's polynomials \cite{grads}, the
 associated normalized eigenvectors (which are the same
as those of $K_\beta$) are
\begin{eqnarray}\label{zkn}
Z^{(k)}_n(E) := \eta_n \sqrt{\frac{\Gamma(n-k+1)}{E \: \Gamma(n+k)}}\:e^{-\beta E} 
\left(2\beta E\right)^k
\mathsf{L}^{(2k-1)}_{n-k}\left(2\beta E\right)\:, \:\:\:\:\mbox{$n=k,k+1,\ldots$,}\label{ZEgen}
\end{eqnarray}
$\eta_n$ being a pure phase which can be 
arbitrarily fixed (in \cite{mopi02,mopi03} we used $\eta_n=1$). As noticed in \cite{mopi02},
if $\beta$ is interpreted as an inverse temperature, the exponential $e^{-\beta E}$ suggests an interpretation
in terms of a canonical ensemble of the energetic content of these states. In this paper we examine in depth this 
possibility finding out very interesting results.\\
If $m=0$ 
and so $\ca{H} \cong L^2(\bR^+,dE)\oplus L^2(\bR^+,dE)$,
an analogue representation exists in each space $L^2(\bR^+,dE)$. Making use of  Heisenberg representation
it is simply proven that the algebra generated by $H,\overline{D},\overline{C}$, with depending-on-time coefficients, is made of
constant of motions \cite{mopi03}. Thus $SL(2,\bR)$ is a {\em symmetry} of the one-particle system. 
That can straightforwardly be extended to the free 
quantum field in Fock space. The crucial point is that the found symmetry is {\em hidden}: 
It cannot be induced by the background
geometry since the Killing fields of Rindler spacetime enjoy a different Lie algebra from that of $H_0,D,C$ 
and no representation of $SL(2,\bR)$ exists in terms of isometries of ${\bf R}$ (see \cite{mopi03}
for definitions and details). The picture changes dramatically when the found $SL(2,\bR)$ symmetry is examined
on the horizon as said in [e] below.

 {\bf [c]} The space {\bf R} is naturally embedded in a Minkowski spacetime which contains the horizon
associated with the rindler metric. 
 Rindler {\em light coordinates} 
$u= t-{\log(\ka y)}/{\ka}$, $v = t+{\log(\ka y)}/{\ka}$ (where $u,v\in \bR$) 
cover the (open) Rindler space  ${\bf R}$. Separately, $v$ is
well defined on the {\em future horizon} ${\bf F}$, $u\to +\infty$, and $u$ is well defined  on the {\em past horizon}
${\bf P}$, $v\to -\infty$ (see figure). A wavefunction in (\ref{wave}) admits well-defined limits 
toward  the future horizon $u\to +\infty$:\begin{eqnarray}
\psi(v)  =  \int \frac{e^{-iEv}}{\sqrt{4\pi E}}
e^{i\rho_{m,\ka}(E)}\tilde\psi_+(E)\: dE + \mbox{c.c.} \label{limitH+}
\end{eqnarray}
$e^{i\rho_{m,\ka}(E)}$ being a pure phase \cite{mopi03}.
In coordinate $u\in \bR$, the restriction of $\psi$ to ${\bf P}$ is similar  with  $v$
replaced for $u$ and  $\rho_{m,\ka}(E)$ replaced by $-\rho_{m,\ka}(E)$. 
If $m=0$ restrictions to ${\bf F}$ and ${\bf P}$ are similar to (\ref{limitH+})with the difference that $e^{i\rho_{m,\ka}(E)}$ is replaced by $1$,
only ingoing  components survive in the limit toward ${\bF}$  and only outgoing  components survive in the limit 
toward ${\bF}$ ($v$ must be replaced for $u$ in that case).
Discarding the phase it is possible to consider the following real
``{\em wavefunction on the (future) horizon ${\bf F}$}'':
\begin{eqnarray}
\varphi(v)  =  \int_{\bR^+} \frac{e^{-iEv}}{\sqrt{4\pi E}}
 \tilde{\varphi}_+(E)\: dE + \int_{\bR^+} \frac{e^{+iEv}}{\sqrt{4\pi E}}
 \overline{\tilde{\varphi}_+(E)}\: dE, \label{fieldonhorizon}
\end{eqnarray}
{\em where $\varphi$ 
 is any real function in Schwartz' space on $\bR\equiv {\bf F}$},
as the basic object in defining a quantum field theory on the future horizon. The same is doable
concerning {\bf P}. 
The space of horizon wavefunctions can be equipped with a
diffeomorphism invariant symplectic form  
$\Omega_{\bf F}(\varphi,\varphi'):= \int_{\bf F} \varphi' d\varphi- \varphi d\varphi'$. A suitable 
causal propagator can also be defined $E_{\bf F}(v,v')= (1/4)sign(v-v')$ and used as said below.
First of all define the Hermitean scalar product $\langle \varphi'_+ , \varphi_+ \rangle_{\bf F} 
:= -i\Omega_{\bf F}(\overline{\varphi'_{+}},\varphi_{+})$.
The one-particle Hilbert space ${\cal H}_{\bf F}$ is the completion with respect to that scalar product
of the space of complex combinations of positive frequency parts $\tilde\varphi_+(E)$,
of horizon wavefunctions $\varphi$.
 As $\langle \varphi'_+ , \varphi_+ \rangle_{\bf F} = \int_{\bR^+} \overline{\tilde{\varphi}'_+(E)} \tilde{\varphi}_+(E)dE$,  ${\cal H}_{\bf F}$
 turns out to be
isomorphic to $L^2(\bR^+,dE)$ once again. 
The field operator is defined in the symmetrized Fock space ${\gF}({\cal H}_{\bf F})$,
with vacuum state $|0\rangle_{\bf F}$,
with rigorous symplectic definition given by
\begin{eqnarray}
 \varphi \mapsto \Omega_{\bf F}(\varphi, \hat\phi_F) &:=& ia(\overline{{\varphi}_+}) -ia^{\dagger}(\varphi_+) \label{PHIF}\:, 
\end{eqnarray}
 where $\varphi$ is any horizon wavefunction in the space specified above.  
 With these definitions, in spite of the 
absence of any 
equation of motion the essential features of free quantum field theory are preserved by that definition \cite{mopi03}.
 Degeneracy of the metric on the horizon prevents from smearing field operators by functions due to the 
 ill-definiteness of the induced volume measure. However, employing the symplectic approach \cite{Wald}, 
 a well-defined smearing-procedure is that of field operators and exact $1$-forms $\eta=df$ where $f=f(v)$, 
 $v\in \bR\equiv {\bf F}$,
 is any real function in Schwartz' space. More precisely, 
$\eta(v) \mapsto E_{\bf F}(\eta) = \frac{1}{4}\int_\bR sign(v-v')\eta(v') = \psi_\eta(v)$
 defines a one-to-one correspondence between exact one-forms and horizon wavefunctions of the form (\ref{fieldonhorizon})
 and $\eta = 2d\psi_\eta$.  Thus, if $\eta= d\varphi$ with $\varphi=\varphi(v)$ in Schwartz' space, one can define
 \begin{eqnarray}
\eta \mapsto \hat\phi_{\bf F}(\eta) := \Omega_{\bf F}(E_{\bf F}\eta, \hat\phi_{\bf F})\:,\label{fsmearedF}
\end{eqnarray}
which is the rigorous meaning of
\beq
\hat{\phi}_{\bf F}(\eta)=\int_0^\infty \frac{dE}{\sqrt{4\pi E}}
\left(\int_\bR e^{-iEv}\eta(v)\right) a_{E}+ 
\left(\int_\bR e^{iEv} \eta(v)\right) a_{E}^\dagger\:.\label{nonrig} 
\eeq
Horizon wavefunctions $\varphi$ and $1$-forms $\eta,\eta'$
in the spaces said above enjoy the same properties as in the bulk. More precisely one has
\begin{eqnarray}\int_{\bf F} \varphi \eta  &=& \Omega_{\bf F}(E_{\bf F}\eta,\varphi) \:\:\mbox{and}\:\:
\int_{\bf R} (E_{\bf F}\eta) \eta' = \Omega_{\bf F}(E_{\bf F}\eta,E_{\bf F}\eta')
\label{psifF}\:, \\
\mbox{$[$}\hat{\phi}_{\bf F}(\eta), \hat{\phi}_{\bf F}(\eta')\mbox{$]$} &=& -i E_{\bf F}(\eta,\eta') = \int_{\bf F} \psi_{\eta'}d\psi_{\eta}- 
  \psi_{\eta} d\psi_{\eta'}\:.
\end{eqnarray}
The latter is nothing but the rigorous meaning of the formal equation
  $[\hat{\phi}(v),\hat{\phi}(v')]=-iE_{\bf F}(v,v')$.  Finally a ``locality property''
 holds true:
  $$[\hat{\phi}_{\bF}(\eta), \hat{\phi}_{\bF}(\eta')]=0\:\:\:\:\:\:
 \mbox{if  $\:\:\:\:supp(\eta) \cap supp(\eta')= \emptyset$.}$$
 Everything we have stated for ${\bf F}$ can analogously be stated
  for ${\bf P}$. 

  {\bf [d]} It is possible to prove the existence of a unitary equivalence between the theory in the bulk and that on the horizon in the sense we
are going to describe.\\ 

\noindent {\em {\bf Theorem 1.1}.
If $f$ is any real smooth compactly supported function $f$
 used to smear the bulk field, define $\eta_f := 2d(E(f)\spa\rest_{\bf F})$,
 and $\omega_f := 2d(E(f)\spa\rest_{\bf P})$,  $E(f)\spa\rest_{{\bf F}/{\bf P}}$
 being the limit toward ${\bf F}$, respectively ${\bf P}$, of $E(f)$ (see figure).\\
{\bf (a)} If $m>0$, there is a unitary map
 $U_{\bf F}: {\gF}({\cal H}) \to {\gF}({\cal H}_{\bf F})$ such that 
$$U_{\bf F} |0\rangle = |0\rangle_{\bf F}\:,\:\:\: \mbox{and}\:\:\:\: U^{-1}_{\bf F} 
 \hat\phi_{\bf F}(\eta_f) U_{\bf F} = \hat\phi(f)\:.$$ 
  {\bf (b)} If $m=0$ two unitary operators arise $V_{{\bf F}/{\bf P}}: {\gF}({\cal H}_{in/out})  \to 
  {\gF}({\cal H}_{{\bf F}/{\bf P}})$ 
 such that $$V_{{\bf F}/{\bf P}} |0\rangle_{in/out} = |0\rangle_{{\bf F}/{\bf P}}$$
 and 
 $$  V^{-1}_{{\bf F}} \hat\phi_{{\bf F}}(\eta_f) V_{{\bf F}} = \hat\phi_{in}(f)\:,\:\:\: \mbox{and}\:\:\:\: 
 V^{-1}_{{\bf P}} \hat\phi_{{\bf P}}(\omega_f) V_{{\bf P}} = \hat\phi_{out}(f)\:.$$ 
 ${\cal H}_{in/out}$ is the bulk Hilbert space
 associated with the ingoing/outgoing modes and $\hat\phi_{in/out}(f)$ is the part of bulk field operator 
 built up using only ingoing/outgoing modes.}  \\

 \noindent Details on the construction of $U_{\bf F}$, $V_{\bf F}$, $V_{\bf P}$ are supplied in \cite{mopi03}. 
  Similarly to the extent in the bulk case, one focuses on the algebra
${\cal A}_{\bf F}$  of linear combinations of product of field operators $\hat\phi_{\bf F}(\omega)$ varying $\omega$
in the space of allowed complex $1$-forms. We assume that ${\cal A}_{\bf F}$ also contains the unit operator $I$. 
The Hermitean elements of ${\cal A}_{\bf F}$ are the natural 
observables associated with the horizon field.
  From an abstract point of view the found algebra is a unital $*$-algebra of formal operators ${\phi}_{\bf F}(\eta)$
with the additional properties $[{\phi}_{\bf F}(\eta), {\phi}_{\bf F}(\eta)] = -i E_{\bf F}(\eta,\eta')$,
${\phi}_{\bf F}(\eta)^*= {\phi}_{\bf F}(\overline{\eta})$ and linearity in the form $\eta$.
({The analogous algebra
of operators in the bulk fulfil the further  requirement $\phi(f)=0$ if (and only if ) $f=Kg$, $K$ being the Klein-Gordon operator.
No analogous requirement makes sense for ${\cal A}_{\bf F}$ since there is no equation of motion on the
horizon.})
${\cal A}_{\bf F}$ can be studied no matter any operator representation in any Fock space. Operator representations
are obtained via GNS theorem once an algebraic state has been fixed \cite{Wald}. ${\cal A}_{\bf P}$ can analogously 
be defined. Below ${\cal A}_{\bf R}$ denotes the unital $*$-algebra associated with the bulk field operator.
If $m=0$, ${\cal A}_{\bf R}$ naturally decomposes as ${\cal A}_{in}\otimes {\cal A}_{out}$ (see \cite{mopi03})
with obvious notation. 
We have the following result which is independent from any choice of vacuum state and Fock representation.
The proof can be found in
\cite{mopi03}.\\

\noindent {\em {\bf Theorem 1.2}. Assume the same notation as in Theorem 1.1 concerning $\eta_f$ and $\omega_f$.\\
{\bf (a)} If $m>0$,  there is a unique 
 injective unital $*$-algebras homomorphism $\chi_{\bf F}: {\cal A}_{\bf R} \to {\cal A}_{\bf F}$ such that 
 $\chi_{\bf F}(\phi(f)) = \phi_{\bf F}(\eta_f)$. Moreover in GNS representations in the respectively associated 
 Fock spaces ${\gF}({\cal H})$, ${\gF}({\cal H}_{\bf F})$ built up over $|0\rangle$ and $|0\rangle_{\bf F}$ 
 respectively, $\chi_{\bf F}$ has a unitary implementation naturally  induced by
   $U_{\bf F}$ (e.g. $\chi_\bF(\hat{\phi}(f)) =  U_\bF \hat{\phi}(f)U^{-1}_\bF$).\\
 {\bf (b)} If $m=0$,   there are two
 injective unital $*$-algebras homomorphisms $\Pi_{{\bf F}/{\bf P}}: {\cal A}_{in/out} \to {\cal A}_{{\bf F}/{\bf P}}$ 
 such that 
 $\Pi_{\bf F}(\phi(f)) = \phi_{\bf F}(\eta_f)$ and $\Pi_{\bf P}(\phi(f)) = \phi_{\bf P}(\omega_f)$ 
 Moreover in GNS representations in the respectively associated 
 Fock spaces ${\gF}({\cal H}_{in/out})$, ${\gF}({\cal H}_{{\bf F}/{\bf P}})$ 
 built up over $|0\rangle_{in/out}$ and $|0\rangle_{{\bf F}/{\bf P}}$ 
 respectively, $\Pi_{\bf F}$ and $\Pi_{\bf P}$ have  unitary implementations and reduce to $V_{\bf F}$
 and $V_{\bf P}$ respectively.}\\

\noindent Notice that, in particular $\chi_{\bf F}$ preserves the causal propagator, in the sense that it must be
$-iE(f,g) = [\phi(f),\phi(g)] = [\phi_{\bf F}(f),\phi_{\bf F}(g)] = -iE_{\bf F}(\eta_f,\eta_g)$.\\

{\bf [e]} Consider quantum field theory 
on ${\bf F}$, but the same result holds concerning ${\bf P}$. In ${\cal H}_{\bf F}\cong L^2(\bR^+,dE)$
define  operators 
$H_{\bF0},D_{\bf F},C_{\bf F}$ as the right-hand side of the equation that respectively defines $H_0,D,C$
in (\ref{ge3F}). They  and their real linear combinations are  essentially self-adjoint if 
restricted to the invariant dense domain ${\cal D}_k^{({\bf F})}$ defined with the same definition as ${\cal D}_k$
in [b]. 
Exactly as in the bulk case,  operators $iH_{\bf F} = i\overline{H_{\bF0}}$ ,$i\overline{D_{\bf F}},i
\overline{C_{\bf F}}$ generate a strongly-continuous unitary $SL(2,\bR)$ 
representation $\{ {U}^{({\bf F})}_g\}_{g\in SL(2,\bR)}$. 
Hence, varying $g\in SL(2,\bR)$, the unitary operators obtained by unitary holography
$ (U_{\bf F}\spa\rest_{\cal H})^{-1} \:{U}^{({\bf F})}_g \:U_{\bf F}\spa\rest_{\cal H}$ define a representation of 
 $SL(2,\bR)$ for the system in the bulk. By construction 
 $(U_{\bf F}\spa\rest_{\cal H})^{-1} \:H_{\bf F} \: U_{\bf F}\spa\rest_{\cal H}= H$. As a consequence
 every $U^{({\bf F})}_{g}$ gives rise to  a $SL(2,\bR)$ symmetry of the bulk field and the group of these symmetries 
 is unitary equivalent to that generated by 
 $iH,i\overline{D},i\overline{C}$. In particular the one-parameter group associated with $H_{\bf F}$ generates $v$-displacements
 of horizon wavefunctions which are equivalent, under unitary holography, to $t$-displacements
 of bulk wavefunctions. Now, it make sense to investigate the {\em geometrical nature} of the $SL(2,\bR)$
 representation $\{ U ^{({\bf F})}_g\}$ that, as we said, induces, up to unitary equivalences, 
 the original $SL(2,\bR)$ symmetry in the bulk, but now can be examined on the horizon. 
 In fact, the symmetry has  a geometrical meaning:
 The action of every ${U}^{({\bf F})}_g$ on a state $\tilde \varphi_+=\tilde\varphi_+(E)$
 is essentially equivalent to the action of a corresponding 
 ${\bf F}$-diffeomorphism on the associated (by (\ref{fieldonhorizon})) horizon wavefunction
 $\varphi$. More precisely  \cite{mopi03}:\\

 \noindent {\em {\bf Theorem 1.3}.  
Assume  $k=1$ in (\ref{ge3F}),  take a matrix $g\in SL(2,\bR)$.  
 Let  $\varphi= \varphi(v)$ be a real Schwartz'  horizon
 wavefunction  with   positive frequency part  $\tilde\varphi_+=\tilde\varphi_+(E)$
 and  such that $\varphi(0)=0$ and $v\mapsto \varphi(1/v)$ belongs to Schwartz' space too. \\
The wavefunction $\varphi_g$ associated with ${U}^{(\bF)}_g\tilde\varphi_+$ reads
 \beq
\varphi_g(v)= \varphi\at \frac {av+b}{cv+d}\ct - \varphi\left(\frac{b}{d}\right) ,\qquad \begin{pmatrix}a&b\\c&d
\end{pmatrix} = g^{-1}.  \label{18}
\eeq
Moreover one has the particular cases:\\
 {\bf (a)} The unitary one-parameter group generated by $iH_{\bf F}$ 
 is associated to the one-parameter group of ${\bf F}$-diffeomorphisms 
 generated by $\partial_v$.\\
 In other words, for every $t\in \bR$ and positive-frequency  part wavefunction $\tilde{\varphi}_+$, the positive-frequency part 
 wavefunction $e^{itH_\bF}\tilde{\varphi}_+$  is associated with the horizon wavefunction
 $\varphi_{g_t}$ such that $$\frac{\partial \varphi_{g_t}}{\partial t}|_{t=0} = - \partial_v \varphi\:.$$
 {\bf (b)}  With the same terminology as in the case (a),  
 the unitary one-parameter group generated by $i\overline{D_{\bf F}}$ is 
 associated to the one-parameter group of ${\bf F}$-diffeomorphisms 
 generated by $v\partial_v$.\\
{\bf (c)}   With the same terminology as in the case (a), 
the unitary one-parameter group generated by $i\overline{C_{\bf F}}$ 
 is associated to the one-group of ${\bf F}$-diffeomorphisms generated by $v^2\partial_v$.}\\

\noindent The term $- \varphi({b}/{d})$ in (\ref{18}) assures that $\varphi_g$ vanishes as $v\to \pm\infty$. Notice 
that the added term disappears when referring to $d\varphi$ rather than $\varphi$.
The group of diffeomorphisms of ${\bf F}\equiv \bR$ used above,
\beq
v\mapsto \frac{av+b}{cv+d} ,\qquad  \begin{pmatrix}a&b\\c&d
\end{pmatrix} \in SL(2,\bR) \label{SL2RF}
\eeq
in fact gives a representation of $SL(2,\bR)$. It  can be obtained by finite composition of one-parameter subgroups associated with
the following three vector fields
on ${\bf F}$:
$\partial_v, v\partial_v, v^2\partial_v$. It is simply proven that the Lie brackets of  $-\partial_v, -v\partial_v, -v^2\partial_v$ produce
the same algebra as the  Lie algebra of $SL(2,\bR)$.
We conclude that the bulk $SL(2,\bR)$-symmetry is manifest when examined on the horizon, in the sense that 
it is induced by the geometry.  
   
\section{From the line to the circle: The full Virasoro Algebra}

\hfill{{\em Noli tangere circulos meos.}}

\hfill{(Archimedes' last words.)}\\

\noindent The algebra of vector fields  $\partial_v, v\partial_v, v^2\partial_v$ 
can be extended to include  the class of fields
defined on the horizon $v^{n+1}\pa_v$ with $n\in \bZ$. It is interesting to notice that these fields (more precisely the fields $-v^{n+1}\pa_v$)
enjoy Virasoro commutation relations  without central charge.
In fact there is a central representation of Virasoro algebra which presents a central charge
and is directly defined in terms of operators acting in the Fock space of the horizon particles.
The representation can be introduced
after one has given a convenient definition of quantum field operator on the circle ${\mathsf{F}} = {\bf F}\cup \{\infty\}$.\\

\noindent {\bf 2.1}. {\em QFT on the circle ${\mathsf{F}} = {\bf F}\cup \{\infty\}$}.
Consider the vector field on ${\bf F}$,
\begin{equation}{\cal K}
:= \frac{1}{2}\left(\beta \partial_v  + \frac{1}{\beta} v^2\partial_v\right)\:.
\label{K}
\end{equation}
That field  is associated with the essentially self-adjoint operator that is defined on ${\cal D}^{(\bF)}_1$
\begin{eqnarray}
K_{{\bf F}\beta} 
:= \frac{1}{2}\left(\beta H_{\bf F} + \frac{1}{\beta}C_{\bf F} \right) \label{Koperator}\:,
\end{eqnarray}
 in the Lie algebra of the unitary representation of $SL(2,\bR)$ because of of Theorem 1.4. It is simply 
proven that the integral line of ${\cal K}$
with origin in $v=0$ is $v=\beta \tan (\theta/2)$ with $\theta\in (-\pi,\pi)$ and $v=0$ corresponding to $\theta = 0$. 
One can use $\theta$ as a new coordinate
on ${\bf F}$ with the advantage that this new coordinate gets finite values in the whole compactfied 
manifold ${\bf F}\cup \{\infty\}\equiv {\mathsf{F}}$ (in the sense of Alexandrov's procedure), the added point $\infty$
corresponding to  $\theta=\pi \equiv -\pi$ in the circle.
As a consequence of our definitions, it turns out that 
\begin{equation}{\cal K}= \partial_\theta\:.
\label{Ktheta}
\end{equation}
In fact this formula smoothly extends the left-hand side on the whole circle  $\sF$.
By construction, there is the natural submanifold embedding ${\bf F}\subset {\mathsf{F}}$. We want to show that
such an inclusion can be extended to free quantum field theory if a suitable definition of QFT on ${\mathsf{F}}$
is given. We follow  a procedure very similar to that
used for the horizon.
As a final result we show that more strongly, the ``inclusion" of QFT on ${\bf F}$ into QFT on ${\mathsf{F}}$
is actually a unitary equivalence as well as a $*$-algebras inclusion. The 
observable $K_{{\bf F}\beta}$ plays a central role in that identification.
The associated quantum field theory on ${\mathsf{F}}$ will be proved to support a nice unitary Virasoro's algebra  representation
with an explicit geometric meaning that extends the unitary representation of $SL(2,\bR)$.\\

\noindent 
Consider the space of real $C^\infty$ functions $\rho$
on ${\mathsf{F}}$, $C^\infty({\mathsf{F}};\bR)$, and define a subsequent real vector space ${\cal S}({\mathsf{F}})$ by taking the quotient with respect 
to the equivalence relation, for $\rho,\rho' \in C^\infty({\mathsf{F}};\bR)$
\begin{eqnarray}
\rho\sim \rho'\:\:\: \mbox{iff} \:\:\: d(\rho-\rho')=0\:. \label{equiv}
\end{eqnarray}
  From now on, the elements of ${\cal S}({\mathsf{F}})$ are called {\em circle wavefunctions}.
The  following symplectic form on ${\cal S}({\mathsf{F}})$ is well-defined and nondegenerate (the latter is not true on 
$C^\infty({\mathsf{F}};\bR)$):
\begin{eqnarray}
\Omega_{{\mathsf{F}}}(\rho,\rho') := \int_{{\mathsf{F}}} \rho' d\rho - \rho d\rho'. 
\end{eqnarray}
The elements of $C^\infty({\mathsf{F}};\bC)$ can be expanded in Fourier series. If $\rho\in C^\infty({\mathsf{F}};\bR)$,
with a re-arrangement of the Fourier coefficients it holds either in $L^2(\sF,d\theta)$ and  in the uniform sense
$$\rho(\theta) = \rho_0+ \sum_{n=1}^\infty \:\:\frac{e^{-in\theta}\:\tilde{\rho}_+(n)}{\sqrt{4\pi n}} +
\frac{e^{in\theta}\:\overline{\tilde{\rho}_+(n)}}{\sqrt{4\pi n}}.$$
As $\tilde{\rho}_+(n)$ is proportional to $\int_{{\mathsf{F}}} e^{-i n\theta}\rho(\theta) d\theta$ and 
$\int_{{\mathsf{F}}} e^{-i n\theta} d\theta =0$,
if $n>0$, the coefficients $\tilde{\rho}(n)$ are not affected if $\rho$ is replaced by $\rho'$ with
$\rho-\rho'=$ constant, and thus the coefficients $\tilde{\rho}_+(n)$, {\em for $n>0$}, are well-associated with
an element of ${\cal S}({\mathsf{F}})$. In the following we indicate the elements of ${\cal S}({\mathsf{F}})$
simply by $\rho$ instead of $[\rho]$. In the sense clarified above, 
if $\rho \in {\cal S}({\mathsf{F}})$ we have the expansion
\begin{eqnarray}
\rho(\theta) = \sum_{n=1}^\infty \:\:\frac{e^{-in\theta}\:\tilde{\rho}_+(n)}{\sqrt{4\pi n}} + c.c. =
\rho_+(\theta) + c.c. .
\label{fourier}
\end{eqnarray}
To define the one-particle Hilbert space,  define the Hermitean scalar product $$\langle \rho'_+ , \rho_+ \rangle_{{\mathsf{F}}} 
:= -i\Omega_{{\mathsf{F}}}(\overline{\rho'_{+}},\rho_{+})\:.$$
The one-particle Hilbert space ${\cal H}_{{\mathsf{F}}}$ is the completion with respect to that scalar product
of the space of complex combinations of positive frequency parts $\{\tilde{\rho}_+(n)\}$,
of circle wavefunctions $\rho$.
It is symply proven that, if $\rho\in C^\infty({\mathsf{F}};\bC)$ with Fourier coefficients  $\{C_n\}_{n\in \bZ}$, for every $p=0,1,\ldots$
there is a real $K_p$ such that $|n|^p |C_n| \leq K_p$  for all $n\in \bZ$. As a consequence,
 $\sum_{n\in \bZ} n |C_n|^2 <\infty$. We conclude that, if $\rho\in {\cal S}({\mathsf{F}})$,
 the sequence of complex numbers $\{\tilde\rho_+(n) = \sqrt{2n}C_n\}_{n=1,2,\ldots}$ is an element of $\ell^2(\bC)$.
A direct computation shows that  ${\cal H}_{{\mathsf{F}}}$ turns out to be 
isomorphic to $\ell^2(\bC)$ because
$$\langle \rho'_+ , \rho_+ \rangle_{{\mathsf{F}}} = \sum_{n=1}^\infty \overline{\tilde{\rho}'_+(n)} \tilde{\rho}_+(n)$$
Using the Hilbert base of ${\cal H}_{\bf F}$
given by the  eigenvectors of 
the operator $K_{\beta \bF}$, $\{Z^{(1)}_n\}_{n=1,2,\ldots}$ 
(where the phase of $Z^{(1)}_n$ in (\ref{zkn}) 
is fixed to be $\eta_n=(-1)^{n+1}$), 
the unitary map $M : {\cal H}_{\bf F} \to {\cal H}_{{\mathsf{F}}}$ can be defined
 such that
\begin{eqnarray}
M :  \varphi \mapsto \{ \langle Z^{(1)}_n, \varphi \rangle\}_{n=1,2,\ldots} 
\label{iso}\:.
\end{eqnarray}
That isomorphism has a natural geometric interpretation stated in the former part of the theorem below.\\

\noindent {\bf Theorem 2.1}. {\em Let $\varphi= \varphi(v)$ be a real horizon wavefunction 
(which belongs to Schwartz' space on $\bR \equiv {\bf F}$)
associated with a quantum state  $\tilde{\varphi}_+$. If
$\rho$ is the circle wavefunction associated with $\varphi$ by means of the unitary 
transformation (\ref{iso}), that is  $\tilde{\rho}_+ :=M(\tilde{\varphi}_+)$, one has
\begin{eqnarray}
\rho(\theta) = \varphi(v(\theta))
\label{expansiontheta}\:,
\end{eqnarray}
where $v(\theta) = \beta \tan(\theta/2)$, $\theta\in (-\pi,\pi]$. In other words
$$\varphi(v(\theta)) = \sum_{n=1}^{\infty} \frac{\langle Z^{(1)}_n, \tilde{\varphi}_+\rangle}{\sqrt{4\pi n}} e^{-in\theta} +c.c.
+\mbox{constant} \:.$$
The linear map $\varphi \mapsto \rho$ defined in (\ref{expansiontheta}) is injective and preserves  the symplectic forms of the respective spaces,
that is, if $\rho'$ is associated with $\varphi'$  by the map (\ref{expansiontheta}) itself,
 \begin{eqnarray}
\Omega_{{\mathsf{F}}}(\rho,\rho') = \Omega_{\bf F}(\varphi,\varphi')
\label{preservation}\:.
\end{eqnarray}}

\noindent {\em Proof.} Notice that, if the real horizon wavefunction $\varphi= \varphi(v)$ is in the
Schwartz' space,
the function $(-\pi,\pi] \ni\theta \mapsto \varphi(v(\theta))$ is well-defined and belongs to $C^\infty({\mathsf{F}};\bR)$
with $\varphi(v(\pm \pi))=0$ with all its derivatives of any order. So the thesis
makes sense.
The second part can straightforwardly be proven by using the given definitions,
so we focus on the former only. 
If the real horizon wavefunction $\varphi= \varphi(v)$ is in the Schwartz' space, the associated
positive frequency part $\tilde{\varphi}_+(E)$ is such that $\tilde{\varphi}_+(E)/\sqrt{4\pi E}$ is the restriction to $\bR^+$ 
of a Schwartz'
function. As a consequence ${\varphi}_+(v) = \int_{0}^{+\infty}dE\, e^{-iEv}\tilde{\varphi}_+(E)/\sqrt{4\pi E}$
is smooth and $$\varphi_+(v(\theta)) \sim \mbox{const.}\:\frac{\tilde{\varphi}_+(E)}{\sqrt{E}}|_{E=0} (\theta\mp \pi)^2$$
as $\theta \to \pm \pi$. So the Fourier expansion of $\varphi_+$ makes sense and each coefficient of the Fourier expansion of $\varphi$
is the sum of the corresponding coefficients of  the Fourier expansion  of $\varphi_+$ and 
 $\overline{\varphi_+}$, it being $\varphi= \varphi_++\overline{\varphi_+}$.
We want to evaluate the Fourier coefficients of $\varphi_+$. First consider the Fourier coefficients with $n>0$.
By direct computation \cite{grads}  one finds
\beq \int_0^{+\infty} \frac{e^{-iE\beta\tan(\theta/2)}Z^{(1)}_n(E)}{\sqrt{4\pi E}}\, dE = \frac{1}{\sqrt{4\pi n}}\left( (-1)^{n+1} + 
e^{-in\theta}\right)\label{X}\:,\eeq
with $\theta\in (-\pi,\pi]$ 
(notice that the dependence form $\beta$ cancels out due the shape (\ref{ZEgen}) of functions $Z^{(1)}_n$ by passing to the new 
variable of integration  $\beta E$
in the integral). As a consequence, defining $Z^{(1)}_n(E)=0$ if $E<0$, inverting the Fourier(-Plancherel) transform (and changing
the integration variable $v\to -v$),
\begin{equation}
\frac{Z^{(1)}_n (E)}{\sqrt{2E}} = 
\lim_{L\to +\infty} \int_{-L}^{L}dv\, e^{-ivE} \frac{(-1)^{n+1} + e^{2in \tan^{-1}(v/\beta)}}{\sqrt{4\pi n}}\:,
\label{inter}\end{equation}
the limit being computed in the sense of $L^2(\bR,dE)$. 
Since $E\mapsto \tilde{\varphi}_+(E)/\sqrt{E}$
is the 
restriction to $\bR^+$ of a Schwartz function, $E\mapsto \psi(E) = \sqrt{2E}\tilde{\varphi}_+(E)$ ($\psi(E) :=0$ for $E<0$ ! ) is
a function in $L^1(\bR, dE)
\cap L^2(\bR, dE)$. The functions $E\mapsto Z^{(1)}_n(E)$ and $E\mapsto Z^{(1)}_n(E)/{\sqrt{2E}}$ (assumed to vanish for $E<0$)
are real and belong to  $L^1(\bR, dE)
\cap L^2(\bR, dE)$. It holds
$$\langle Z^{(1)}_n, \tilde{\varphi}_+\rangle  = \int_0^\infty \overline{Z^{(1)}_n(E)}\tilde{\varphi}_+(E) dE =
\int_{-\infty}^\infty \frac{Z^{(1)}_n(E)}{\sqrt{2E}}\psi(E) dE\:.$$
Using (\ref{inter}) and
taking the $L^2$-continuty of the scalar product into account,
one gets
$$\langle Z^{(1)}_n, \tilde{\varphi}_+\rangle = \lim_{L\to +\infty}
\int_{-\infty}^{\infty} dE\, \psi(E) \int^{L}_{-L} dv\,  e^{-ivE} \frac{(-1)^{n+1} + e^{2in \tan^{-1}(v/\beta)}}{\sqrt{4\pi n}}\:,
$$
that is
$$\langle Z^{(1)}_n, \tilde{\varphi}_+\rangle = \lim_{L\to +\infty}
\int_{0}^{\infty} dE\, \tilde{\varphi}_+(E) \int^{L}_{-L} dv\, 2E 
\frac{e^{-ivE}}{\sqrt{2E}} \frac{(-1)^{n+1} + e^{in \theta(v)}}{\sqrt{4\pi n}}\:,
$$
Using $E e^{-ivE} = i\frac{\partial}{\partial v} e^{-ivE}$ and integrating by parts it arises
$$\langle Z^{(1)}_n, \tilde{\varphi}_+\rangle = C(L) +\lim_{L\to +\infty} \sqrt{2n} 
\int_{0}^{\infty} dE \tilde{\varphi}_+(E) \int^{L}_{-L} dv \frac{e^{-iEv}}{\sqrt{4\pi E}}
\frac{e^{in\theta(v)}}{\sqrt{2\pi}}\frac{d\theta}{dv}\:,$$
where $C(L)$ is a boundary term which vanishes in the limit $L\to \infty$ by Riemann-Lebesgue's lemma. Interchanging the integration
symbols and taking the limit as $L\to \infty$ we finally get the $n$-th Fourier coefficient of $\varphi_+$ and 
the $(-n)$-th Fourier coefficient of $\overline{\varphi_+}$
\begin{eqnarray}
 \int_{-\pi}^{\pi} \varphi_+(v(\theta))\frac{e^{in\theta}}{\sqrt{2\pi}} \:d\theta
= \frac{\langle Z^{(1)}_n, \tilde{\varphi}_+\rangle}{\sqrt{2n}}\:\:\:\:,\:\:\:\:
\int_{-\pi}^{\pi} \overline{\varphi_+(v(\theta))}\frac{e^{-in\theta}}{\sqrt{2\pi}} \:d\theta
= \frac{\overline{\langle Z^{(1)}_n, \tilde{\varphi}_+\rangle}}{\sqrt{2n}} \:. \nonumber
\end{eqnarray}
Now we pass to consider the remaining Fourier coefficients.
Since in (\ref{inter}) $Z^{(1)}_n(E)$ is defined to vanish for $E<0$,
one has that, if $n>0$
$$
\lim_{L\to +\infty} \int_{-\infty}^{0} dE\, f(E) \int_{-L}^{L}dv\, e^{-ivE} 
\frac{(-1)^{n+1} + e^{2in \tan^{-1}(v/\beta)}}{\sqrt{4\pi n}} =0 \:,$$
which, after complex conjugation and change of variables $E\to -E$, is equivalent to 
\begin{equation}
\lim_{L\to +\infty} \int_{0}^{+\infty} dE\, g(E) \int_{-L}^{L}dv\, e^{-ivE} 
\frac{(-1)^{m+1} + e^{2im \tan^{-1}(v/\beta)}}{\sqrt{4\pi |m|}} =0 \:,
\label{inter2}\end{equation}
where $m= -n <0$ and $g\in L^2(\bR,dE)$. Using $g(E)= \sqrt{2E}\tilde{\varphi}_+(E)$ for $E\geq 0$ and $g(E)=0$ otherwise
and following the same procedure as for the case $n>0$, (\ref{inter2}) implies that, if $n<0$
$$ \int_{-\pi}^{\pi} \varphi_+(v(\theta))\frac{e^{in\theta}}{\sqrt{2\pi}} \:d\theta
= 0\:.
$$
As a consequence,
$$\int_{-\pi}^{\pi} \overline{\varphi_+(v(\theta))}\frac{e^{-in\theta}}{\sqrt{2\pi}} \:d\theta
= 0\:.
$$
Putting all together we get
$$\varphi(v(\theta)) =\varphi_+(v(\theta)) +\overline{\varphi_+(v(\theta))} =
\mbox{constant} + \sum_{n=1}^{\infty} \frac{\langle Z^{(1)}_n, \tilde{\varphi}_+\rangle}{\sqrt{4\pi n}} e^{-in\theta} 
+\sum_{n=1}^{\infty} \frac{\overline{\langle Z^{(1)}_n, \tilde{\varphi}_+\rangle}}{\sqrt{4\pi n}} e^{in\theta} 
$$
which concludes the proof. $\Box$\\

\noindent The result stated in Theorem 2.1 suggests to define a quantum field on the circle ${\mathsf{F}} := {\bf F}\cup \{\infty\}$
whose Hilbert space is the symmetrized Fock space ${\gF}({\cal H}_{{\mathsf{F}}}) \cong  {\gF}({\cal H}_{\bf F})$,
where the isomorphism is that naturally induced by $M$ of eq. (\ref{iso}) and the vacuum $|0\rangle_{{\mathsf{F}}}$ is associated with
 $|0\rangle_{\bf F}$ by the isomorphism itself.
Formally the quantum field
operator on ${\mathsf{F}}$ reads
\begin{eqnarray}
\hat{\phi}(\theta) = \sum_{n=1}^\infty \:\:\frac{e^{-in\theta}\:\alpha_n}{\sqrt{4\pi n}} + \frac{e^{in\theta}\:\alpha^\dagger_n}{\sqrt{4\pi n}}
\label{formal}\:,
\end{eqnarray}
where $\alpha_n$ and $\alpha^\dagger_n$ are the annihilator and constructor operator of modes $Z^{(1)}_n$.\\
The field operator is defined in the symmetrized Fock space ${\gF}({\cal H}_{{\mathsf{F}}})$,
with rigorous symplectic definition given by
\begin{eqnarray}
 \rho \mapsto \Omega_{{\mathsf{F}}}(\rho, \hat\phi_F) &:=& i\alpha(\overline{{\rho}_+}) -i\alpha^{\dagger}(\rho_+) \label{CHIF}\:,
\end{eqnarray}
 where $\rho$ is any circle wavefunction and respectively $\alpha(\overline{{\rho}_+})$,
 $\alpha^{\dagger}(\rho_+)$ annihilates and creates the states $\overline{\rho_+}$ and  $\rho_+$.
 Once again a well-defined smearing-procedure is that of field operators and exact $1$-forms $\eta=d\rho$ where $\rho \in {\cal
 S}({\mathsf{F}})$. Notice that $d\rho$ does not depend on the chosen element of the class of equivalence associated with $\rho$.
  More precisely,  we introduce the ``{\em causal propagator}'' on ${\mathsf{F}}$
$$\eta(\theta) \mapsto E_{{\mathsf{F}}}(\eta) = \frac{1}{4}\int_{{\mathsf{F}}} [sign(\theta-\theta') + (\theta'-\theta)/\pi]\eta(\theta')\:,$$
where it is understood that one has to take the quotient with respect the equivalence relation
   defining ${\cal S}(\sF)$  after the action  of $E_{{\mathsf{F}}}$.
$E_{{\mathsf{F}}}$ gives rise to  a bijective linear  map from the space of
 real exact $C^\infty$  one-forms  on ${\mathsf{F}}$
   (which will be denoted by ${\cal D}({\mathsf{F}})$) and  ${\cal S}({\mathsf{F}})$ itself. 
   Indeed, it results that if $\rho\in {\cal S}({\mathsf{F}})$
\beq
E_{{\mathsf{F}}}(\omega) = \rho \:\:\:\:\:\mbox{if and only if}\:\:\:\: \omega=2d\rho\:. \label{useful}
\eeq
We can define the field operator smeared by elements of ${\cal D}({\mathsf{F}};\bR)$ as
 \begin{eqnarray}
\eta \mapsto \hat\phi_{{\mathsf{F}}}(\eta) := \Omega_{{\mathsf{F}}}(E_{{\mathsf{F}}}\eta, \hat\phi_{{\mathsf{F}}})\:.\label{fsmearedS1}
\end{eqnarray}
which is the rigorous meaning of
\beq
\hat{\phi}_{{\mathsf{F}}}(\eta)=\sum_{n=1}^\infty 
\left(\int_{{\mathsf{F}}} e^{-in\theta}\eta(\theta)\right) \frac{\alpha_{n}}{\sqrt{4\pi n}}+ 
\left(\int_{{\mathsf{F}}} e^{in\theta} \eta(\theta)\right) \frac{\alpha_{n}^\dagger}{\sqrt{4\pi n}}\:. 
\eeq
Circle wavefunctions $\rho$ and $1$-forms $\eta,\eta'$
in the spaces said above enjoy the same properties as in the bulk. More precisely one has
\begin{eqnarray}\int_{{\mathsf{F}}} \rho \eta  &=& \Omega_{{\mathsf{F}}}(E_{{\mathsf{F}}}\eta,\rho) \:\:\mbox{and}\:\:
\int_{{\mathsf{F}}} (E_{{\mathsf{F}}}\eta) \eta' = \Omega_{{\mathsf{F}}}(E_{{\mathsf{F}}}\eta,E_{{\mathsf{F}}}\eta')
\label{psifS1}\:, \\
\mbox{$[$}\hat{\phi}_{{\mathsf{F}}}(\eta), \hat{\phi}_{{\mathsf{F}}}(\eta')\mbox{$]$} &=& -i E_{{\mathsf{F}}}(\eta,\eta') \label{useful0} \:.
\end{eqnarray}
The latter is nothing but the rigorous meaning of the formal equation
  $[\hat{\phi}_{{\mathsf{F}}}(\theta),\hat{\phi}_{{\mathsf{F}}}(\theta')]=-iE_{{\mathsf{F}}}(\theta,\theta')$.\\
 Notice that as a consequence of (\ref{useful}), (\ref{psifS1}), (\ref{useful0}),
  a ``locality property''
 holds
 $$[\hat{\phi}_{{\mathsf{F}}}(\eta), \hat{\phi}_{{\mathsf{F}}}(\eta')]=0\:\:\:\:\:\:
 \mbox{if  $\:\:\:\:supp(\eta) \cap supp(\eta')= \emptyset$.}$$

 Everything we said about the future circle ${\mathsf{F}} = {\bf F}\cup \{\infty\}$ can be restated, with 
 obvious changes of notation, for the past circle  ${\mathsf{P}} := {\bf P}\cup \{\infty\}$. \\
 Theorem 2.1 together with Theorems 1.1 and 1.2 has two straightforward consequences.\\

\noindent {\em {\bf Theorem 2.2}. 
If $f$ is any real smooth compactly supported function $f$
 used to smear the bulk field, extend on $\mathsf{F}$ and $\mathsf{P}$ the forms 
 $\eta_f$ and $\omega_f$ defined in Theorem 1.1 by putting $\eta_f(\infty):=0$ and $\omega_f(\infty):=0$ and consider these forms as elements of ${\cal D}(\sF)$ and ${\cal D}(\sP)$
   respectively.\\
{\bf (a)} If $m>0$, there is a unitary map
 $U_{\mathsf{F}}: {\gF}({\cal H}) \to {\gF}({\cal H}_{\mathsf{F}})$ such that 
$$U_{\mathsf{F}} |0\rangle = |0\rangle_{\mathsf{F}}\:,\:\:\: \mbox{and}\:\:\:\: U^{-1}_{\mathsf{F}} 
 \hat\phi_{\mathsf{F}}(\eta_f) U_{\mathsf{F}} = \hat\phi(f)\:.$$ 
  {\bf (b)} If $m=0$, two unitary operators arise $V_{{\mathsf{F}}/{\mathsf{P}}}: {\gF}({\cal H}_{in/out})  \to 
  {\gF}({\cal H}_{{\mathsf{F}}/{\mathsf{P}}})$ 
 such that $$V_{{\mathsf{F}}/{\mathsf{P}}} |0\rangle_{in/out} = |0\rangle_{{\mathsf{F}}/{\mathsf{P}}}$$
 and 
 $$  V^{-1}_{\mathsf{F}} \hat\phi_{\mathsf{F}}(\eta_f) V_{\mathsf{F}} = \hat\phi_{in}(f)\:,\:\:\: \mbox{and}\:\:\:\: 
 V^{-1}_{\mathsf{P}} \hat\phi_{\mathsf{P}}(\omega_f) V_{\mathsf{P}} = \hat\phi_{out}(f)\:.$$ 
 ${\cal H}_{in/out}$ is the bulk Hilbert space
 associated with the ingoing/outgoing modes and $\hat\phi_{in/out}(f)$ is the part of bulk field operator 
 built up using only ingoing/outgoing modes.}  \\

\noindent {\em Sketch of proof}. The unitary operator $U_{\bf F}$ in Theorem 1.1 is obtained (see \cite{mopi03}) as the
unitary operator that fulfils the following pair of conditions. (1) $U_{\bf F}|0\rangle = |0\rangle_{\bf F}$;
 (2) for every natural $n$, consider the 
subspace of ${\gF}({\cal H})$,
${\cal H}^{n\otimes}$, spanned by (symmetrized) states with $n$ particles; on every ${\cal H}^{n\otimes}$,
$U_{\bf F}$ reduces  to the tensor product of $n$ copies of the unitary operator 
${\cal U}_{\bf F}: {\cal H} \to {\cal H}_{\bf F}$,
where, under the identifications (working in the energy representations) 
${\cal H}\cong L^2(\bR^+,dE)$, ${\cal H}_{\bf F}\cong L^2(\bR^+,dE)$, ${\cal U}_{\bf F}$ is nothing but
 the identity operator.
Now consider the composite unitary 
operator ${\cal U}_{{\mathsf{F}}}:=  M \circ {\cal U}_{\bf F} : {\cal H} \to {\cal H}_{{\mathsf{F}}}$, where
$M$ is as in eq. (\ref{iso}), and
define $U_{{\mathsf{F}}}$ such that $U_{{\mathsf{F}}} |0\rangle = |0\rangle_{{\mathsf{F}}}$ and the restriction of $U_{{\mathsf{F}}}$
to every ${\cal H}^{n\otimes}$ coincides with to the tensor product of $n$ copies of the unitary operator 
${\cal U}_{{\mathsf{F}}}$. Theorems 1.1 and 2.1 and the definition of $\hat{\phi}_{{\mathsf{F}}}$ immediately imply 
the validity of the thesis. The case of $m=0$ can be proven by the same way.
$\Box$\\

 \noindent Similarly to the extent on the horizon case, one can focus on the algebra
${\cal A}_{\bf {\mathsf{F}}}$  of linear combinations of product of field operators $\hat\phi_{{\mathsf{F}}}(\omega)$, varying $\omega$
in the space ${\cal D}({\mathsf{F}};\bC):= {\cal D}({\mathsf{F}}) + i {\cal D}({\mathsf{F}})$
and defining $\hat\phi_{{\mathsf{F}}}(\omega + i \omega'):= \hat\phi_{{\mathsf{F}}}(\omega)+ i \hat\phi_{{\mathsf{F}}}(\omega')$. 
We assume that ${\cal A}_{{\mathsf{F}}}$ also contains the unit operator $I$. 
The Hermitean elements of ${\cal A}_{{\mathsf{F}}}$ are the natural 
observables associated with the horizon field.
 From an abstract point of view the found algebra is a unital $*$-algebra of formal operators ${\phi}_{{\mathsf{F}}}(\eta)$
with the additional properties $[{\phi}_{{\mathsf{F}}}(\eta), {\phi}_{{\mathsf{F}}}(\eta')] = -i E_{{\mathsf{F}}}(\eta,\eta')$,
${\phi}_{{\mathsf{F}}}(\eta)^*= {\phi}_{{\mathsf{F}}}(\overline{\eta})$ and linearity in the form $\eta$.
${\cal A}_{{\mathsf{F}}}$ can be studied no matter any operator representation in any Fock space. Operator representations
are obtained via GNS theorem once an algebraic state has been fixed \cite{Wald}. Everything we said can be
extended to the analogous $*$-algebra defined on $\mathsf{P}$, ${\cal A}_{\mathsf{P}}$. We have a second result.\\

\noindent   \noindent {\em {\bf Theorem 2.3}. Assume the same notation as in Theorem 2.2
concerning $\eta_f$ and $\omega_f$.\\
{\bf (a)} If $m>0$,  there is a unique
 injective unital $*$-algebras homomorphism $\chi_{\sF}: {\cal A}_{\bf R} \to {\cal A}_{\sF}$ such that
 $\chi_{\sF}(\phi(f)) = \phi_{\sF}(\eta_f)$. Consider the GNS representations in the 
 Fock spaces ${\gF}({\cal H})$, ${\gF}({\cal H}_{\sF})$ built up over $|0\rangle$ and $|0\rangle_{\sF}$
 respectively associated with ${\cal A}_{\bf R}$ and  ${\cal A}_{\sF}$,in these representations  $\chi_{\sF}$ has a unitary implementation naturally induced by $U_{\sF}$
   (e.g. $\chi_{\sF}(\hat{\phi}(f)) = U_\sF\hat{\phi}(f)U^{-1}_\sF$).\\
 {\bf (b)} If $m=0$,   there are two
 injective unital $*$-algebras homomorphisms $\Pi_{{\sF}/{\sP}}: {\cal A}_{in/out} \to {\cal A}_{{\sF}/{\sP}}$
 such that
 $\Pi_{\sF}(\phi(f)) = \phi_{\sF}(\eta_f)$ and $\Pi_{\sP}(\phi(f)) = \phi_{\sP}(\omega_f)$.
 Moreover, considering the  GNS representations in the 
 Fock spaces ${\gF}({\cal H}_{in/out})$, ${\gF}({\cal H}_{{\sF}/{\sP}})$
 built up over $|0\rangle_{in/out}$ and $|0\rangle_{{\sF}/{\sP}}$
 respectively associated with ${\cal A}_{in/out}$ and ${\cal A}_{{\sF}/{\sP}}$, $\Pi_{\sF}$ and $\Pi_{\sP}$ have  unitary implementations and reduce to $V_{\sF}$
 and $V_{\sP}$ respectively.}\\

 \noindent {\em Sketch of proof}. Consider the map $\chi'_{\sF} : \hat{\phi}_{\bf F}(\eta) \mapsto 
 \hat{\phi}_{\sF}(\eta)$ where $\eta = d\varphi$, $\varphi$ being any real Schwartz function on ${\bf F}\equiv \bR$.
 In $\hat{\phi}_{\bf F}(\eta)$, $\eta$ is supposed extended to the whole $\sF$ by means of  $\eta(\infty):=0$ so that 
 $\eta \in {\cal D}(\sF;\bR)$. Using the fact that it holds $[\hat{\phi}_{\sF}(\eta), \hat{\phi}_{\sF}(\eta')] = -i
 \Omega_{\sF}(E_{\sF}(\eta'),E_{\sF}(\eta)) = -i
 \Omega_{\bf F}(E_{\bf F}(\eta'),E_{\bf F}(\eta)) =
 [\hat{\phi}_{\bf F}(\eta), \hat{\phi}_{\bf F}(\eta')]$, one proves that $\chi'_{\sF} $ 
 uniquely extends into a injective
 $*$-algebra homomorphism from ${\cal A}_{\bf F}$ to ${\cal A}_{\sF}$. The  injective $*$-algebra homomorphism
 $\chi_\sF$ is nothing but $\chi'_{\sF}\circ \chi_{\bf F}$. The remaining properties are strightforward 
 consequences of the properties of $\chi_{\bf F}$ stated in Theorem 1.2. The case $m=0$ is analogous.
$\Box$\\

\noindent {\bf 2.2}. {\em Virasoro algebra with $c=1$ in the Fock space of the circle.} 
 The unitary  map  $M : {\cal H}_{\bf F} \to {\cal H}_{\sF}$ associates 
the essentially self-adjoint operators  $H_{\bF0}, D_{\bf F},C_{\bf F}$ defined on ${\cal D}_1^{(\bF)}\subset {\cal H}_{\bf F}$
with analogous essentially self-adjoint  operators acting on one-particle  circle states ${\cal H}_{\sF}$,  respectively  $H_{\sF0}, D_{\sF}, C_{\sF}$.  
More precisely, the  real linear combinations of these operators are  essentially self-adjoint
in the dense invariant  domain ${\cal D}_1^{(\sF)} = M({\cal D}^{(\bF)}_{1})$ spanned by 
the eigenvectors of  $K_{\beta {\sF}}$  associated with the analogous operator $K_{\beta {\bF}}$
(\ref{Koperator}).
The Lie algebra spanned by the  operators above in $\sF$  gives rises to a strongly-continuous unitary $SL(2,\bR)$ 
representation $\{U^{(\sF)}_g\}_{g\in SL(2,\bR)}$ on the Hilbert space of the circle 
that is related, by means of $M$, 
with the analogous unitary representation $\{U^{{(\bf F})}_g\}_{g\in SL(2,\bR)}$
found on the horizon ${\bf F}$ discussed in  [e]
of 1.2.  And thus,  in turn,  it induces 
just the bulk symmetry induced by  $\{U^{{(\bf F})}_g\}_{g\in SL(2,\bR)}$ (see [e] of 1.2)
by means of  unitary holography.
In particular $H_{\sF} := \overline{H_{\sF 0}}$ turns out to be associated with the generator of Rindler-time displacements $H$.\\
     $H_{\sF0}, D_{\sF}, C_{\sF}$ are a basis of the Lie algebra
   of $SL(2,\bR)$.
An equivalent, but more useful in the following,  basis of the Lie algebra of $SL(2,\bR)$
made of  essentially self-adjoint operators in ${\cal D}^{(\sF)}_1$ is that of the operators  $K_{\beta \sF}, D_{\sF}, S_\sF$ with
 \beq
S_{\sF}:= \frac{1}{2}\left(\beta H_{\sF0} - \frac{1}{\beta}C_{\sF}\right)\label{Sf} \:.\eeq
 Now, it make sense to investigate the {\em geometrical nature} of the $SL(2,\bR)$
 representation $\{{U}^{(\sF)}_g\}$ on {\em the circle $\sF$ instead of the horizon ${\bf F}$}.
 First of all one has to notice that the vector fields 
 $\frac{\partial}{\partial v}, v\frac{\partial}{\partial v}, v^2\frac{\partial}{\partial v}$, which
 give rise to the geometric interpretation of $\{{U}^{(\bF)}_g\}$ when working on ${\bf F}$,
  span in ${\bf F}$  the same space as that spanned by  the three smooth vector fields 
defined  on the whole circle $\sF$, 
$$\partial_\theta\:, \cos(\theta) {\partial_\theta}\:, \sin(\theta) {\partial_\theta}\:,$$
 The proof  is straightforward using the relation $v= \beta \tan (\theta/2)$ only.   
Then  consider  the transformations (\ref{SL2RF}) of the line ${\bf F}$, translate 
them  in the variable $\theta = 2 \arctan (v/\beta)$ extended to the domain  $(-\pi,\pi]$ so to
include $\infty$. The new transformations  so obtained define a  representation of $SL(2,\bR)$  in terms of  orientation-preserving
diffeomorphisms $d_g$
of the circle ${\sF}$: 
\beq
d_g : \theta  \mapsto2\arctan \left(\frac {a\beta \tan(\theta/2)+b}{c\beta^2 \tan (\theta/ 2)+\beta d}\right),
 \qquad  g:= 
\begin{pmatrix}a&b\\c&d \end{pmatrix} \in SL(2,\bR)\:.
\eeq
There is another, more elegant,  way  to write the elements of  same diffeomorphism group:
\beq
\delta_{h} : e^{i \theta}
  \mapsto
 \frac{\zeta e^{i\theta} +\overline{\eta}}{\eta e^{i\theta} +\overline{\zeta}},
 \qquad  h:= 
\begin{pmatrix}\zeta &\overline{\eta} \\{\eta}&\overline{\zeta} \end{pmatrix} \in SU(1,1)\:,
\label{hm}
\eeq
where we have used the group isomorphism $SL(2,\bR)\ni g \mapsto h \in SU(1,1)$ with:
$$\zeta:= \frac{\beta a+\beta d +i(b-\beta^2 c)}{2}\:,\:\:\: \eta:= \frac{\beta d-\beta a-i(b+\beta^2
c)}{2}\:.$$
The condition $h\in SU(1,1)$, when $h$ has the form  in (\ref{hm}), can equivalently be 
written
\beq
|\zeta|^2-|\eta|^2 =1 \label{hm'}\:.
\eeq

\noindent{\em Remark}. Notice that the transformation 
$\rho \mapsto \rho_g$ does not mix Fourier components with positive frequency
and Fourier components with negative frequency and {\em vice versa}. (This fact allows one 
to look for unitary representations of the considered group in the one-particle Hilbert space
which is constructed by using positive frequency part of wavefunctions.)
Indeed, using (\ref{hm}),  $e^{-in\theta}$ is mapped into $\left( \frac{\zeta e^{i\theta}+\overline{\eta}}{\eta e^{i\theta}+\overline{\zeta}}\right)^{-n}$.
Fourier coefficients with strictly  ``negative frequency'' $-m$ are proportional to the integrals, where
$m,n \geq 1$ are integer,
$$\int_{-\pi}^\pi \left( \frac{\zeta e^{i\theta}+\overline{\eta}}{\eta e^{i\theta}+\overline{\zeta}}\right)^{-n}
e^{-im\theta} d\theta =  \oint_{+\bS^1}  \frac{-i}{z^{m+1}}
\left(\frac{\eta z + \overline{\zeta}}{\zeta z+ \overline{\eta}}\right)^n dz =  \oint_{+\bS^1}  \frac{-i}{z^{m+1}}
\left(\frac{\eta}{\zeta}+\frac{1/\zeta}{\zeta z+ \overline{\eta}}\right)^n dz\:,$$
where $+\bS^1$ is the circle $|z|=1$ with positive orientation.
Expanding the expression under the last integration symbol using the binomial formula one reduces 
to a linear combination of contributions of integrals with form
$$ \oint_{+\bS^1} \frac{dz}{z^{m+1} (z- z_0)^p}$$
with $p=0,1,\ldots n\geq 1$, $m\geq 1$ and where $z_0= -\overline{\eta}/\zeta$ with $\zeta\neq 0$ (due to (\ref{hm'})). 
If $p>0$, using the condition (\ref{hm'}) one sees that, whatever the values of $\eta$ and $\zeta$
and barring the pole at $z=0$
with order $m+1$, there is  another pole of order $p$
inside the region with boundary $\bS^1$, at $z= -\overline{\eta}/\zeta$. Cauchy formula for $p>0$ proves that the contribution of the 
two residues in each integral cancel out each other
and the final result is zero. The case $p=0$ gives the same result automatically. So it make sense
to look for unitary representation of the group in the one-particle space.\\
  
\noindent  Exactly as the case of Theorem 1.3, if 
 $\rho$ is a real circle wavefunction, 
with associated one-particle quantum state $\tilde \rho_+=\tilde\rho_+(n)$,
the action of every ${U}^{(\sF)}_g$ on $\tilde{\rho}_+$
is equivalent to the action of a corresponding 
 ${\sF}$-diffeomorphism,  $d_g$, on the horizon wavefunction $\rho$ itself.\\
   
\noindent  {\em {\bf Theorem 2.4}. Assume $k=1$ in the definition of ${\cal D}_k^{(\sF)}$, that is, in  (\ref{ge3F}).  If  $g\in SL(2,\bR)$ and  
$\tilde\rho_+=\tilde\rho_+(n)$ is
 the positive frequency part  of $\rho\in {\cal S}(\sF)$,
the state ${U}^{(\sF)}_g\tilde\rho_+$ can be associated with 
 the wavefunction $\rho_g\in {\cal S}(\sF)$ with
\beq
\rho_g(\theta) = \rho(d^{-1}_g \theta) \:\:,\:\:\:\:\:\mbox{for all $\theta\in (-\pi,\pi]$} \label{grap}.
\eeq
In particular (with the same terminology as that used in (a) of Theorem 1.3):\\
{\bf (a)}  The unitary one-parameter group generated by $i\overline{K_{{\sF}\beta}}$ 
 is associated with the one-group of ${\sF}$-diffeomorphisms 
 generated by $\partial_\theta$;\\
{\bf (b)} the unitary one-parameter group generated by $i\overline{D_{\sF}}$ is 
 associated to the one-parameter group of ${\sF}$-diffeomorphisms 
 generated by $\sin \theta \partial_\theta$;\\
{\bf (c)}  the unitary one-parameter group generated by $i\overline{S_\sF}$ 
 is associated to the one-group of ${\sF}$-diffeomorphisms generated by $\cos \theta \partial_\theta$.\\
The Lie algebra spanned by   fields $\partial_\theta$, $\cos\theta \partial_\theta$, $\sin\theta \partial_\theta$ is a realization of the
Lie algebra of $SL(2,\bR)$.}\\

\noindent{\em Proof}. 
The first part can be proven as follows. Take $\rho\in C^\infty(\sF;\bR)$.
 As $ {\cal H}_{\sF}= \ell^2(\bC)$, 
 the associated positive frequency part in ``frequency picture''  $\tilde{\rho}_+\in {\cal H}_{\sF}$ is a sequence 
 $\{C_n\}_{n=1,2,\ldots}$. 
The associated positive frequency part in ``$\theta$ picture'' can be written 
 $$\rho_+(\theta) = {\cal F}(W \tilde{\rho}_+)(\theta) $$
 where ${\cal F}: \ell^2(\bC) \to L^2((-\pi,\pi), d\theta)$ 
 and $W : \ell^2(\bC) \to \ell^2(\bC)$ are respectively
  the {\em continuous} linear operators, 
  \begin{eqnarray}
  {\cal F} &:&     \{C_n\}_{n=1,2,\ldots} \mapsto \sum_{n=1}^{+\infty}
 C_n \frac{e^{-in\theta}}{\sqrt{2\pi}}  \:,\\
 W &:& \{C_n\}_{n=1,2,\ldots} \mapsto \left\{\frac{C_n}{\sqrt{2n}}\right\}_{n=1,2,\ldots}\:.
 \end{eqnarray}
 On the other hand, in the sense of the topology of $\ell^2(\bC)$,
  $$\tilde{\rho}_+= \sum_{m=1}^\infty C_m \Psi_m$$
 where $\{\Psi_m\}_{m=1,2,\ldots}$ is the Hilbert base $\Psi_m = \{\delta_{mn}\}_{n=1,2,\ldots}$.
By linearity, continuity and the absence of positive-negative frequency mixing,
\beq (\rho_g)_+ = {\cal F}(W U^{(\sF)}_g \tilde{\rho}_{+}) = {\cal F}\left(W U^{(\sF)}_g \sum_nC_n \Psi_n\right) 
= \sum_n  {\cal F}\left(W U^{(\sF)}_g C_n \Psi_n\right) 
\:.\label{ZZ}\eeq
Now one notice that Theorem 1.3 holds true also for the  horizon wavefunction ${\varphi}^{(n)}$ 
with positive frequency part, in ``frequency picture'' given by 
$Z_n^{(1)}(E)$
(with  and $k=1,2,\ldots$) as the associated positive frequency wavefunction. (The proof of this fact is, in
fact, the same proof as that of Theorem 1.3, that is Theorem 4.7 in \cite{mopi03},  with trivial adaptations
which make sense in the considered case. In particular Proposition 4.2 in \cite{mopi03}
can  directly be
proven by using (\ref{X}). It is useful to notice that  $\Theta Z_n^{(1)} = (-1)^n Z_n^{(1)}$.)
By (\ref{X}),
\beq
\varphi^{(n)}(v(\theta))= 
\frac{1}{\sqrt{4\pi n}}\left[ (-1)^{n+1}(C_n+\overline{C_n}) + 
C_ne^{-in\theta}+ \overline{C_n}e^{in\theta}\right]\:,
\label{Y'0}\eeq
and thus, Theorem 1.3 says that
\beq \varphi^{(n)}_{g}(v(\theta))= 
\frac{1}{\sqrt{4\pi n}}\left[ 
C_ne^{-ind^{-1}_g\theta} - e^{-ind^{-1}_g\pi}C_n \right] +c.c.\:. 
\label{Y'}\eeq
Since, by Theorem 2.1, $Z_n$ in  ${\cal H}_{\bf F}$ is transformed into  $\Psi_n$ of ${\cal H}_\sF$
and the same transformation associates $U^{(\bF)}_g$ with $U^{(\sF)}_g$,
(\ref{Y'}) can be re-written in the space ${\cal H}_{\sF}$ and in the circle $\sF$:
\beq {\cal F}\left(W U^{(\sF)}_g  C_n\Psi_n\right)(\theta) + c.c. = C_n\varphi_g^{(n)}(v(\theta)) + c.c. =
\frac{1}{\sqrt{4\pi n}}\left[ 
C_ne^{-ind^{-1}_g\theta}- C_ne^{-ind^{-1}_g \pi}
\right] +c.c.
\:. \label{Z}  \eeq
Inserting it in (\ref{ZZ}), one concludes that
\beq \rho_g(\theta) =  
\sum_{n=1}^{+\infty} 
C_n \frac{e^{-ind^{-1}_g \theta}}{\sqrt{4\pi n}} - C_n \frac{e^{-ind^{-1}_g \pi}}{\sqrt{4\pi n}} +c.c.\:.
\eeq
The convergence must be understood  in the sense of $L^2(\sF,d\theta)$. However 
since $\rho \in C^\infty(\sF)$ and thus $\rho\circ d^{-1}_g\in C^{\infty}(\sF)$,
the latter admit a uniformly convergent Fourier series and the series
$\sum_n {|C_n|}/{\sqrt{4\pi n}}$ converges too. By 
the uniqueness property of Fourier series it must hold,
$$\rho_g(\theta) = \rho(d^{-1}_g \theta) + constant \:\:,\:\:\:\mbox{for all $\theta\in \sF$}\:.$$
This concludes the proof if the  functions are considered as elements of ${\cal S}(\sF)$.\\
Let us pass to prove the statement in (b), the remaining cases can be proven following a
 strongly analogous proof (which is much more simple in the case (a)).
 If $t \mapsto g_t$ is the one-parameter subgroup of  $SL(2,\bR)$
 whose associated one-parameter group of diffeomorphisms is that generated by
 the field $\sin\theta\: \partial_\theta$, consider the transformed wavefunction
 $\rho_{g(t)}(\theta):= \rho(d^{-1}_t(\theta))=  \rho(d_{-t}(\theta))$ 
 and the associated positive frequency part $\widetilde{\rho_{g(t)}}_+$ in ``frequency representation''.
 Since $SL(2,\bR)$ acts on positive frequency wavefunctions by means of a  strongly continuous unitary representation,
  there must be some self-adjoint  generator $A$  (not depending on $\rho_+$)
  such that $\widetilde{\rho_{g(t)}}_+ = e^{itA}\tilde{\rho}_+$. Our thesis
  states that $A=\overline{D_\sF}$.  With the formalism introduced above, the statement
  turns out to be proved  if it holds for all  the states $\widetilde{\rho^{(k)}}_+= \Psi_k = \{\delta_{nk}\}_{n=1,2,\ldots}\in \ell^2(\bR)$,
   $k=1,2,\ldots$. Hence we want to show that for $k=1,2,\ldots$,
   $\widetilde{\rho^{(k)}_{g(t)}}_+= e^{it\overline{D_\sF}}\Psi_k$.  
  To this end it is sufficient to show  that 
   in the topology of ${\cal H}_{\sF}\cong \ell^2(\bC)$,
 \beq \left.\frac{d}{dt} \widetilde{\rho^{(k)}_{g(t)}}_+ \right|_{t=0} = iD_\sF \Psi_k\:. \label{stone}
 \eeq 
  Indeed,  by  Stone's theorem the derivative in the left-hand side is $iA\Psi_k$, on the other hand,
   since  $D_\sF$ is essentially self-adjoint in the
  linear space finitely spanned by the vectors $\Psi_k$, it  must be $A=\overline{D_{\sF}}$.  
  Let us prove (\ref{stone}). From now on $\widetilde{\rho^{(k)}_{g(t)}}_+ =\{C_n(t)\}_{n=1,2,\ldots}$.   Defining 
  $\theta_t(\theta):= d_{-t}(\theta)$, one has 
  \beq C_n(t) = \sqrt{\frac{n}{k}} \int_{-\pi}^{\pi} \frac{e^{-i(k\theta_t(\theta)- n\theta)}}{2\pi}
  d\theta\:,\label{ct}\eeq
because of (\ref{Y'}) where, by construction, $\rho^{(k)}_{g(t)}(\theta) = \varphi^{(k)}_{g(t)}(v(\theta))$ and using the fact that there is no mixing of 
positive end negative frequencies under the action of the group.  
By direct computation one finds that $C_n(0)=0$ and $dC_n(t)/dt|_{t=0}=0$ if $n\neq k,k\pm 1$.
  So if the derivative in the left-hand side of (\ref{stone}) is computed term by term, (\ref{stone})
  can be re-written
  \beq \frac{ \delta_{n,k-1} \sqrt{k(k-1)}-  \delta_{n,k+1} \sqrt{k(k+1)}}{2} = \langle \Psi_n, iD_{\sF} \Psi_k\rangle_{\sF}
  \label{ncf}\eeq
  where we have computed the derivatives of $dC_n(t)/dt|_{t=0}$ using (\ref{ct}).
    However, it also holds 
    $$\langle \Psi_n, iD_{\sF} \Psi_k\rangle_{\sF} =\langle  Z^{(1)}_n, iD_{\bF} Z^{(1)}_k\rangle_{\bF}\:,$$
    and the right-hand side can be computed trivially (for instance by employing the formalism in p. 137 of \cite{mopi02})
  and it turns out to coincide with the left-hand side of (\ref{ncf}), so (\ref{ncf}) holds true.   \\
  To conclude the proof,   it is sufficient to show that 
   the derivative in left-had side of (\ref{stone}),
  which is computed with respect to the topology of $\ell^2(\bC)$, can equivalently be computed
  deriving term by term the sequence of complex which defines $\widetilde{\rho^{(k)}_{g(t)}}_+$.
  Expanding the term under the integral symbol in (\ref{ct}) by means of Taylor formula in the variable $t$ about $t=0$,
  using the Lagrange formula for the remnant and, finally, using integration by parts and
   the fact that the integrated functions are smooth and periodic on $\bS^1$, one proves that
  for some constant $A$, for all $t$ in a neighborhood of $0$
  and for all $n\neq k, k\pm 1$:
  \beq \left|\frac{C_n(t)}{t}\right|^2 \leq \frac{A}{n^2}\label{mag}\:. \eeq
    Thus $$\sum_{n=1}^{\infty} \left| \frac{C_n(t)-C_n(0)}{t} - \frac{dC_n(t)}{dt}|_{t=0}\right|^2 =
  \sum_{n\neq k,k\pm 1} \left| \frac{C_n(t)}{t} \right|^2 + \sum_{n=k,k\pm 1} \left| \frac{C_n(t)-C_n(0)}{t} - \frac{dC_n(t)}{dt}|_{t=0}\right|^2  \:.$$
  $\frac{C_n(t)}{t} \to 0$ in our hypotheses for $0< n\neq k, k\pm 1$ and so the sum of the corresponding series above vanishes too
   due to Lebesgue's dominated convergence
theorem
  with respect to the Dirac measure with support on the points $n\neq k, k\pm 1$
  as a consequence of (\ref{mag}). We finally gets
  $$\lim_{t\to 0}\sum_{n=1}^{\infty} \left| \frac{C_n(t)-C_n(0)}{t} - \frac{dC_n(t)}{dt}|_{t=0}\right|^2 =
  \lim_{t\to 0}\sum_{n=k, k \pm 1} \left| \frac{C_n(t)-C_n(0)}{t} - \frac{dC_n(t)}{dt}|_{t=0}\right|^2=0  \:.$$   
  We conclude that the derivative in the left-hand side of (\ref{stone}) computed with respect to the
  topology of $\ell(\bC)$ coincides with that computed term by term. This concludes the proof because
the last statement can straightforwardly be proven by direct inspection.  
    $\Box$ \\

\noindent The theorem states that, in fact,   the  bulk $SL(2,\bR)$ symmetry
becomes manifest when examined on the circle $\sF= \bF \cup \{\infty\}$.
 However that is not the whole story because the found circle unitary $SL(2,\bR)$
 representation is just a little part of a larger unitary  representation with 
 geometrical meaning.  We can, in fact, consider the Lie algebra 
$Vect(\bS^1)$  of the infinite dimensional Lie group \cite{loopgroupbook}
of orientation-preserving  diffeomorphisms
of the circle $Diff^+(\bS^1)$, where $\bS^1=\sF$ in our case.
To make contact with Virasoro algebra we have to consider an associated {\em complex} Lie algebra \cite{KR}.  Consider  the complex Lie
algebra   $ d(\sF) :=  Vect(\sF)\oplus i Vect(\sF)$
equipped with usual Lie brackets  $\{\cdot, \cdot\}$ and the involution  $\omega : X \mapsto -\overline{X}$ for $X\in d(\sF)$,
so that  $\omega(\{X,Y \}) = \{\omega(Y),\omega(X)\}$.
An algebraic  basis of that algebra is made of 
the complex  smooth fields on $\sF$:
\begin{equation}
{\cal F}_{n} := i e^{in\theta} \partial_\theta\:,\:\:\:\:\:\:\: \mbox{with $n\in \bZ$.} \label{F_n}
\end{equation}
 The vector fields ${\cal F}_n$ enjoy the celebrated {\em Virasoro commutation rules} with central charge $c=0$:
\beq\{{\cal F}_{n},{\cal F}_{m} \} = (n-m) {\cal F}_{ n+m}\:, \label{virc=0}\eeq
and the Hermiticity condition
\beq \omega({\cal F}_n) = {\cal F}_{-n}\:. \eeq
In the presented picture $Vect(\sF)$ is nothing but  the sub-algebra of $d(\sF)$ 
containing all of the vectors 
fixed under $-\omega$. An algebraic basis of $Vect(\sF)$ is that made of the fields
 \beq    {\cal F}^{(+)}_n := \frac{\omega({\cal F}_{n}) + {\cal F}_n}{2i} 
=\cos(n\theta) \partial_\theta\:\:\:, \:\:\:
 {\cal F}^{(-)}_m := \frac{\omega({\cal F}_{m}) - {\cal F}_m}{2}  = \sin(n\theta) \partial_\theta\:, 
\label{virwrong}\eeq
where $n=0,1,\ldots$ while $m=1,2,\ldots$. Conversely, the base of $d(\sF)$, $\{{\cal F}_n\}_{n\in \bZ}$ can be obtained from  the base above
as, where $n =1,2,\ldots$,
 \beq    {\cal F}_0 :=i {\cal F}^{(+)}_0 \:, \qquad
 {\cal F}_n :=  {\cal F}_n^{(+)} + \: i {\cal F}_n^{(-)} \:, \qquad
 {\cal F}_{-n} :=  {\cal F}_{-n}^{(+)} - \: i {\cal F}_{-n}^{(-)}\:. \eeq
Notice that the three fields ${\cal F}^{(+)}_0,  {\cal F}^{(+)}_1, {\cal F}^{(-)}_1$ are in fact generators of a finite-dimensional sub algebra
of $Vect(\sF)$, namely the
representation of the Lie algebra of $SL(2,\bR)$
found above, which is equivalently generated by the three fields $v^{n+1}\partial_v$ with
$n=-1,0,1$. However for $|n| >1$, the algebras spanned by generators $v^{n+1}\partial_v$ and ${\cal F}^{(\pm)}_n$
are different and we focus attention on the latter ones.  \\
 By direct inspection one proves that each of the  fields  ${\cal F}^{(\pm)}_n\in Vect(\sF)$ generate a {\em global}  one-parameter
group of $\sF$ orientation-preserving diffeomorphisms (this fact does not hold for fields $v^{n+1}\partial_v$ in $\bF$). Global means here
that the additive parameter which labels the group ranges over the entire real line $\bR$.
 In turn, that group of diffeomorphisms 
generates  a group of automorphisms of the algebra of
the quantum field ${\cal A}_\sF$. Let us explain how it happens. 
 If  $d^{({\cal F}^{(\pm)}_n)}_\lambda : \sF\to \sF$ is an element of the one-parameter 
 (orientation-preserving) diffeomorphism group generated by ${\cal F}^{(\pm)}_n$, $\lambda\in \bR$ being the addictive parameter, 
 and $\rho \in {\cal S}(\sF)$, as usual we define the associated group of wavefunction transformations:
 $(\alpha_\lambda^{({\cal F}^{(\pm)}_n)}\rho)(\theta) := \rho(d^{({\cal F}^{(\pm)}_n)}_{-\lambda}(\theta))$.
 Notice that the  transformations $\alpha_\lambda^{({\cal F}^{(\pm)}_n)}$ 
are, in fact, 
 automorphisms of the real vector space ${\cal S}(\sF)$ equipped with the symplectic 
 form $\Omega_{\sF}$ because the latter is orientation-preserving diffemorphism invariant.\\
 
\noindent {\em Remark}.  By direct inspection  one realizes that, if $n>1$ the transformation $\alpha_\lambda^{({\cal F}^{(\pm)}_n)}$ 
does {\em not} admit the space of positive frequency wavefunctions as invariant space. As a consequence it is not 
possible to represent $\alpha_\lambda^{({\cal F}^{(\pm)}_n)}$ unitarily 
in the one-particle space ${\cal H}_\sF$. To implement the transformation $\alpha_\lambda^{({\cal F}^{(\pm)}_n)}$ 
at quantum level, the entire Fock space is necessary if $n>1$.\\

\noindent As we want to deal with quantum fields smeared by exact $1$-forms
 of ${\cal D}(\sF)$, we define a natural action of the diffeomorphisms  $d^{({\cal F}^{(\pm)}_n)}_\lambda$
 also on these forms by using (\ref{useful}):
  If $\omega\in {\cal D}(\sF)$ and $\rho: = E_\sF(\omega)$,
 we define the one-parameter group of transformations of $1$-forms 
 $\left\{\beta^{({\cal F}^{(\pm)}_n)}_\lambda\right\}_{\lambda\in \bR}$,
 such that  \beq \beta_\lambda^{({\cal F}^{(\pm)}_n)}(\omega) := 2d\alpha_\lambda^{({\cal F}^{(\pm)}_n)} \left(E_{\sF} (\omega)\right)\:.\label{defbeta}\eeq
 Finally we can define the action of quantum fields by means of
  \beq
  \gamma^{({\cal F}^{(\pm)}_n)}_\lambda (\hat\phi_{\sF}(\omega)) := 
\hat\phi_{\sF}\left(\beta_{-\lambda}^{({\cal F}^{(\pm)}_n)}(\omega) \right)\:. \label{defgamma}\eeq
Using the given definitions, the  fact that $\alpha_\lambda^{({\cal F}^{(\pm)}_n)}$
preserves $\Omega_\sF$ as well as (\ref{psifS1}) and (\ref{useful0}), one finds
 \beq 
 \left[  \gamma^{({\cal F}^{(\pm)}_n)}_\lambda (\hat\phi_{\sF}(\omega)) , \:\:
 \gamma^{({\cal F}^{(\pm)}_n)}_\lambda (\hat\phi_{\sF}(\omega')) \right] =
 [\hat\phi_{\sF}(\omega), \hat\phi_{\sF}(\omega')]\:. \eeq
It is possible to prove by that identity that $\gamma^{({\cal F}^{(\pm)}_n)}_\lambda$
naturally extends into a $*$-algebra automorphism of the algebra ${\cal A}_{\sF}$.
The procedure to do it is very similar to those used in \cite{mopi03} to extend 
transformations of field operators into   $*$-algebra homomorphisms.
So, in fact, every field ${\cal F}^{(\pm)}_n$ gives rise to a one-parameter group
of automorphisms of the algebra ${\cal A}_{\sF}$ that we indicate by
$\left\{  \gamma^{({\cal F}^{(\pm)}_n)}_\lambda\right\}_{\lambda\in \bR}$ once again.
A natural question arises: \\

\noindent {\em Is there a  representation of the (infinite dimensional complex) Lie algebra 
$d(\sF)$, in terms of  operators defined in $\gF({\cal H}_{\sF})$ such that
the fields ${\cal F}^{(\pm)}_n$ are mapped into  (essentially) anti-self-adjoint  operators 
$-iF^{(\pm)}_n$, whose associated unitary one-parameter groups 
implement the respective one-parameter group of ${\cal A}_{\sF}$-automorphisms 
$\left\{  \gamma^{({\cal F}^{(\pm)}_n)}_\lambda\right\}_{\lambda\in \bR}$ (at least at the first order)?  That is 
\beq
e^{-i\lambda \overline{F^{(\pm)}_n}}  \overline{\hat\phi_{\sF}(\omega)}e^{i\lambda \overline{F^{(\pm)}_n}} =
\overline{\gamma^{({\cal F}^{(\pm)}_n)}_\lambda(\hat\phi_{\sF}(\omega))}\label{center}\:,\eeq
 or some other formally related, but  perhaps weaker, identity holds.}\\

\noindent  The answer is yes provided one uses  an operator algebra
which represents a {\em central extensions} of the algebra $d(\sF)$.
In other words one has to permit to change, at quantum level, the relation (\ref{virc=0})
by adding in  the right hand side a  further term which commutes with the elements of the  representation itself. The obtained algebra  is properly
called {\em Virasoro's algebra}.\\
 More precisely,  the  quantum representation on a hand is a straightforward extension of
that previously found for the group $SL(2,\bR)$, on the other hand  is, in fact, a {\em positive-energy} and  {\em unitary} representation  of Virasoro
algebra
with {\em central charge $c=1$} \cite{KR}. To built up such a representation the entire Fock space, and not only the one-particle Hilbert space,
is necessary.
A relevant  point is that the found unitary representation of the Virasoro algebra can be exported in the bulk via unitary
holography.  \\
In the circle Fock space $\gF({\cal H}_{\sF})$, consider the basis obtained by taking all the
symmetrized tensor products of one-particle states $Z^{(1)}_n$, namely the eigenvectors of
the operator $K_{\sF \beta}$.  Henceforth
$\alpha_n$ and $\alpha^\dagger_n$ are respectively
the creation and annihilation operator associated with
the one-particle state $Z^{(1)}_n$ with $n=1,2,\ldots$.  As a consequence
\beq
[\alpha_n,\alpha^\dagger_m] = \delta_{n,m}I\:,\qquad
[\alpha_n,\alpha_m] =[\alpha^\dagger_n,\alpha^\dagger_m] = 0\:. \label{CCR}\eeq
Now,  fix $\mu\in \bR$ and introduce the operators, $a_n$, with $n\in \bZ$ such that
\begin{eqnarray}
a_n=  \begin{cases} \:\:\:\: \mu I &\mbox{if $n=0$}\:,\\ 
 \:\: \: i\sqrt{n}\:\alpha_n &\mbox{if  $n>0$}\:,
\\ -i\sqrt{-n}\:\alpha^{\dagger}_{-n} &\mbox{if $n<0$} \end{cases}\:.\label{an}
\end{eqnarray}
By (\ref{CCR}) 
these operators satisfy the {\em oscillator algebra} commutation relations \cite{KR}
\beq
[a_m, a_n] =m \delta_{m,-n}I
 \label{OCR}\:,\eeq
 and the so called {\em Hermiticity conditions}
 \beq
 a_n^\dagger = a_{-n} \label{hermiticity}
 \eeq
 (actually in the left-hand side is considered only the restriction of $a_n^\dagger$ to the domain of
$a_{-n}$).
With these definitions, the formal expression for $\hat{\phi}_{\sF}$ (\ref{formal})
takes the form
\beq
\hat{\phi}_{\sF}(\theta) = \frac{1}{i\sqrt{4\pi}}\sum_{n\in \bZ\setminus \{0\}} \frac{e^{-in\theta}}{n} a_n \label{formal2}\:,
\eeq
moreover, formally {\em but also with a rigorous meaning in terms of a field operator smeared by an exact $1$-form
} (\ref{fsmearedS1}), it holds
\beq
a_n =\frac{1}{\sqrt{\pi}} \int_{\sF} \hat{\phi}_{\sF}(\theta)\: de^{in\theta}\:, \:\:\:\:\mbox{if $n\in \bZ\setminus \{0\}$} \label{formal3}\:.
\eeq
 Finally, 
 define the operators (denoted by $L_k$ in \cite{KR})
\beq
F_k   &:=&  \frac{\epsilon_k}{2} a^2_{k/2} + \sum_{n>-k/2}a_ {-n} a_{n+k}\:,\:\:\:\:\:
k\in \bZ \label{Fk}
\eeq
where $\epsilon_k=0$ if $k$ is odd, $\epsilon_k=1$ if $k$ is even (including $k=0$).  The various sums are, in fact,  finite when acting on a vector,
 since we adopt  as a common domain  of those operators,  the dense 
subspace  ${\gD}_1^{(\sF)}\subset \gF({\cal H}_{\sF})$  made of  the finite linear combination of  vectors containing any finite number of
particles in states $Z^{(1)}_n$. \\

\noindent {\bf Theorem 2.5}.
{\em The operators $F_k$, $k\in \bZ$ enjoy the following properties on their domain ${\gD}_1^{(\sF)}$.\\
{\bf (a)} The complex Lie algebra  finitely spanned  by operators $F_k$
(equipped with the usual operator commutator and Hermitean conjugation) is
a positive-energy unitary  Virasoro  algebra representation $Vir(\sF)$
with central charge $c=1$.  Indeed it holds
\beq
[F_m,F_n] = (m-n) F_{m+n}+ \delta_{m,-n}\frac{m^3-m}{12}I \:,\:\:\:\:\: \mbox{for $n,m\in \bZ$} \label{VC=1}\:,
\eeq
 Hermiticity relations are fulfilled
\beq
F_m^\dagger \Psi  = F_{-m} \Psi \:,\:\:\:\: \mbox{for every $\Psi \in {\gD}_1^{(\sF)}$}\:, \label{hermiticityL}\eeq
$F_0$ is essentially self-adjoint and $\overline{F_0}$ is positive defined with 
discrete spectrum \beq
\sigma(\overline{F_0})= \left\{\left.\frac{\mu^2}{2}+ N \:\:\right|\:\:N=0,1,\ldots\right\} \label{spec}\:. \eeq 
{\bf (b)} For $n=1,2,\ldots$, the operators 
\beq F^{(+)}_0 := F_0\:,\:\:\: F^{(+)}_n := \frac{F_{-n}+F_{n}}{2} \:\:\:\: \mbox{and}\:\:\:\: F_n^{(-)} := i\frac{F_{-n}-F_{n}}{2} 
\label{CnSn}\eeq are essentially self-adjoint in ${\gD}_1^{(\sF)}$.
({It is worth stressing that the interplay of 
 fields ${\cal F}^{(\pm)}_n$ and ${\cal F}_n$  is the same as  that operators 
 $-iF^{(\pm)}_n$ and $F_n$ and {\em not} $F^{(\pm)}_n$ and $F_n$, this is because
the operator involution $^\dagger$ corresponds to the field involution $\omega$ instead of the simpler
complex conjugation.})}\\

\noindent {\em Proof}. Barring  the statements on  ${F_0}$, the properties in (a) are proven in \cite{KR} 
(see sections 2.1, 2.2, 2.3 and 3.1) as consequences of 
(\ref{OCR}), (\ref{hermiticity}) and (\ref{Fk}). 
The operators $F_0$, $F_m+F_{-m}$ and $i(F_m-F_{-m})$ 
are symmetric by construction and one can prove by direct inspection that the element of ${\gD}_1^{(\sF)}$
are analytic vectors for these operators. Thus they are essentially self-adjoint.
This proves (b) and the essential self-adjointness of $F_0$.
By direct inspection and using (\ref{an}),  one finds
\beq F_0 = \frac{\mu^2}{2}I + \sum_{n=1}^\infty n \alpha^\dagger_n\alpha_n\:. \label{ff0}\eeq
The Hilbert basis of $\gF({\cal H}_\sF)$, $\{|L\rangle \}_{L\in \bN}$ made of the vectors 
(labeled with an arbitrary order by the index $L$) 
containing any finite number of states
$Z_m^{(1)}$ 
is a basis of eigenvectors of $F_0$.  The eigenvalue associated with $|L\rangle$  is of the form
$\mu^2/2 + N_L$ where $N_L$ ranges everywhere in $\bN$.  
This fact suggests to consider  the self-adjoint operator
$$ F_0':=\sum_{L=0}^\infty N_L |L\rangle \langle L|$$
where now  the sum is  interpreted in the strong operator topology in the domain ${\cal D}(F_0')$
containing the vectors $|\Psi\rangle  \in {\gF}({\cal H}_{\sF})$ with 
$$\sum_{l=0}^\infty N_L^2 |\langle L| \Psi \rangle |^2 <\infty\:.$$
By construction, $F_0'$  has the spectrum (\ref{spec}). On the other hand, since 
$F_0 \subset F_0'$ by construction, uniqueness of self-adjoint extensions of $F_0$ implies
$\overline{F_0}=F'_0$.
$\Box$ \\

\noindent  Finally we show that (1) the whole Virasoro representation 
extends the circle $SL(2,\bR)$ unitary representation and (2) it  has the geometric 
meaning (\ref{center}).\\

\noindent {\bf Theorem 2.6}. {\em Referring to the Virasoro representation of Theorem 2.5,\\
{\bf (a)} if (and only if) $\mu=0$, the operators $F^{(+)}_{0},F_1^{(+)},F_1^{(-)}$ admit ${\cal D}^{(\sF)}_1\subset  {\cal H}_{\sF}$ as invariant space
and 
\begin{eqnarray}
F^{(+)}_0\spa\rest_{{\cal H}_\sF} &=& K_{\beta\sF}\:,\\
F^{(+)}_1\spa\rest_{{\cal H}_\sF} &=& S_\sF\:, \label{ni1}\\
F^{(-)}_1\spa\rest_{{\cal H}_\sF} &=& D_{\sF} \:, \label{ni2}
\end{eqnarray}
and so these operators generate the $SL(2,\bR)$ representation $\{U^{(\sF)}_{g}\}_{g\in SL(2,\bR)}$.\\
{\bf (b)}  If (and only if) $\mu=0$, for every  $n\in \bN$ ($n>0$ in the case $^{(-)}$),
(\ref{center}) holds true at the first order at least, 
\beq  \left[F_n^{(\pm)}, \hat{\phi}_\sF(\omega)   \right] = i\left.\frac{d}{d\lambda}\right|_{\lambda=0} \gamma_\lambda^{({\cal F}_n^{(\pm)})}\left(\hat{\phi}_\sF(\omega) \right)\label{weakcenter}
\eeq
for every $\omega = \eta +i \eta' \in {\cal D}(\sF;\bC)$ such that the real wavefunctions $E_{\sF}\eta$ and $E_{\sF}\eta'$ 
are associated with states in  ${\cal D}^{(\sF)}$ and the derivative is computed in the strong operator topology in ${\gD}^{(\sF)}_1$.}\\
 
 \noindent {\em Proof}.  (a) The proof of the first case is a trivial consequence 
of (\ref{ff0}).  Concerning the second and third cases, we notice that using operators 
 $\alpha_n$ and $\alpha^\dagger_n$,
 \begin{eqnarray}
F_{-1} &=&  -i\mu \alpha^\dagger_1 
 + \sum_{n=1}^{\infty} \sqrt{n(n+1)} \alpha^\dagger_{n+1}\alpha_n  \label{mo1}\\
 F_1 &=& i\mu \alpha_1
 +  \sum_{n=1}^{\infty} \sqrt{n(n+1)} \alpha^\dagger_{n}\alpha_{n+1} \label{mo2}\:. 
 \end{eqnarray}
It is obvious that, because of the terms containing $\mu (\alpha^\dagger_1 \pm \alpha_1)$,
the operators above admit ${\cal D}^{(\sF)}_1$ as an invariant space if and only if $\mu=0$.
 In that case, the restrictions to ${\cal D}^{(\sF)}_1$ coincide respectively with the operators 
 $A_+$ and $A_-$ defined in \cite{DFF} or  (23) of \cite{mopi02} 
 (where the coefficient $\beta$ is indicated by
 $\lambda/\kappa$ and $k=1$). With our notations
 \beq A_\pm = \frac{1}{2} \left(\beta H_{\sF 0} - \frac{1}{\beta}C_{\sF}\right) \mp iD_{\sF} \label{Apm}\:,\eeq
 so that $A_-Z^{(1)}_1=0$ and $A_+Z^{(1)}_n = \sqrt{n(n+1)} Z^{(1)}_{n+1}$.
 (\ref{Apm}) implies (\ref{ni1}) and (\ref{ni2}) straightforwardly  taking (\ref{Sf}) into account.\\
Let us come to the last part.   It is symply proven that every exact $1$-form $\omega = \eta + i\eta'$, where the real exact $1$-forms
$\eta,\eta'$ determine circle wavefunctions with positive frequency in ${\cal D}^{(\sF)}_1$, is a finite complex linear combination of forms
 $\omega_m(\theta) := d e^{i m \theta}$
with  $m\in \bZ\setminus \{0\}$ .  Hence it is sufficient
to prove (\ref{weakcenter}) for every $\hat{\phi}(\omega_m)$ with $m\in \bZ\setminus \{0\}$.
Fix $m\in \bZ\setminus\{0\}$ and  $k\in \bZ$. By  direct computation and using (\ref{formal3}) 
and (2.12) in \cite{KR}, one finds
that, for every $\Psi \in \gD^{(1)}_\sF$,
\beq [L_k^{(\pm)}, \hat{\phi}(\omega_m)]\Psi = -\frac{m\sqrt{\pi}}{2}(i)_{\pm} (a_{m-k}\pm a_{m+k})\Psi
\label{Q} \eeq
where $(i)_j := 1$ if $j=+$,  $(i)_j := i$ if $j=-$. {\em The identity above holds provided $a_{m-m}$ and $a_{-m+m}$ are interpreted as
the multiplicative operator $\mu I$}
\cite{KR}.
 On the other hand,
$$\gamma_\lambda^{({\cal F}_k^{(\pm)})}\left(\hat{\phi}_\sF(\omega_m)\right) = i\alpha(\phi_{m\lambda}) - i\alpha^\dagger(\psi_{m\lambda})\:,$$
where the vectors $\psi_{m\lambda}$ and $\phi_{m\lambda}$ are defined by
\beq \psi_{m\lambda} := \left\{\sqrt{n} \int_{-\pi}^\pi e^{in\theta} e^{im\theta_t(\theta)}d\theta \right\}_{n=1,2,\ldots}\:,\:\:\:\:\:\:
\phi_{m\lambda} := \left\{\sqrt{n} \int_{-\pi}^\pi e^{-in\theta} e^{im\theta_t(\theta)}d \theta\right\}_{n=1,2,\ldots} \label{phipsi} \eeq
and $\lambda \mapsto \theta_\lambda(\theta)$ is the integral curve of 
${\cal F}^{(\pm)}_k$ starting from $\theta$.  Notice that the linear maps $\psi \mapsto \alpha(\psi)\Psi$
and $\psi \mapsto \alpha^\dagger(\psi)\Psi$ are  continuous for every fixed vector $\Psi\in \gD^{(1)}_{\sF}$, so that
$$\frac{d}{d\lambda}|_{\lambda=0}   \gamma_\lambda^{({\cal F}_k^{(\pm)})}  \left(\hat{\phi}_\sF(\omega_m)\right)\Psi  =
i\alpha\left(\frac{d}{d\lambda}|_{\lambda=0}\phi_{m\lambda}\right)\Psi - i\alpha^\dagger\left(\frac{d}{d\lambda}|_{\lambda=0}\psi_{m\lambda}\right)\Psi\:.
$$
In turn, using a procedure very similar to that used in the proof of (b) in Theorem 2.4,
one sees that the derivatives $\frac{d}{d\lambda}|_{\lambda=0}\phi_{m\lambda}$
and $\frac{d}{d\lambda}|_{\lambda=0}\psi_{m\lambda}$ evaluated by using the topology of $\ell^2(\bC)$ coincide with the analogous derivatives
computed term-by-term for the
sequences of $\ell^2(\bC)$ which define $ \phi_{m\lambda}$
and $ \psi_{m\lambda}$ (\ref{phipsi}).  These derivatives can be computed straightforwardly 
and give rise to
$$\frac{d}{d\lambda}|_{\lambda=0}    \gamma_\lambda^{({\cal F}_k^{(\pm)})} \left(\hat{\phi}_\sF(\omega_m)\right)\Psi =
-\frac{im\sqrt{\pi}}{2}(i)_{\pm} (a_{m-k}\pm a_{m+k})\Psi\:,$$
{\em where, in the right-hand side, $a_{m-m}$ and $a_{-m+m}$ must be interpreted as the null
operator}.
By comparison with (\ref{Q}) we find that  (\ref{weakcenter}) holds true provided $\mu=0$.
$\Box$\\

\noindent {\em Remarks}.{\bf (1)} A natural question concerns whether or not ${\gF}({\cal H}_\sF)$  is irreducible with respect to the found
Virasoro representation.
The answer depend on the value of $\mu$. If and only if  $\mu\in \bZ$ (and in particular if $\mu=0$) the answer is negative because of
several results by Kac, Segal and Wakimoto-Yamada (see Theorem 6.2 in \cite{KR} where the parameter $l$ used
below is indicated by $\mu$ which differs from the parameter $\mu$ used herein).  If $\mu=-m\in \bZ$, one has
the orthogonal decomposition $${\gF}({\cal H}_\sF) = \bigoplus_{k\in \bZ^+, k\geq -m} V(1, (m+2k)^2/4)\:,$$
 where $V(c, h)$ is the  up-to-isomorphism  unique highest-weight unitary Virasoro representation (which is irreducible by consequence) 
 with central charge $c$ and weight $h$.  We recall the reader that if $c=1$ and $h=l^2/4$ with $l\in \bZ$, $V(c, h)$
 is not a {\em Verma representation}. In other words  the system of generators of $V(c,h)$ 
built up over the singular vector of $V(c,h)$ by application of products of Virasoro
generators 
 contains  linearly dependent vectors. Conversely, if $c=1$ and $h\neq l^2/4$ with $l\in \bZ$,  $V(c, h)$ is a Verma representation.\\ 
 {\bf (2)} It is possible to build up a free scalar   standard $2D$-CFT by using $\hat\phi_{\sF}$
and the analogous field $\hat\phi_{\mathsf{P}}$
defined on ${\mathsf{P}}:= \bP \cup \{\infty\}$. In fact, consider the Wick rotation in Rindler coordinates $t \mapsto it$. Under that
continuation, light-Rindler coordinates transforms into $v\to it + \log( \kappa y)/\kappa $, $u\to it - \log( \kappa y)/\kappa$
and so $\theta = 2\arctan (v/\beta) \to z$, $\theta' = 2\arctan (u/\beta) \to \overline{z}$.  $\theta'$ is the coordinate
on ${\mathsf{P}}$ which is defined analogously to $\theta$. With the given definitions,
$z$ turns  out to be defined on a cylinder $\mathsf{C}$ obtained by taking $Im(z)\in \bR$ and $Re(z)\in (-\pi,\pi]$ with the identification
$-\pi\equiv \pi$. By this way the fields $\hat{\phi}_{\sF}$ and $\hat{\phi}_{\mathsf{P}}$ become respectively the Euclidean {\em holomorphic} and
{\em anti holomorphic} fields in $\gF({\cal H}_\sF)\otimes \gF({\cal H}_{\mathsf{P}})$:
$$ \hat{\phi}(z) = \sum_{n\in \bZ\setminus \{0\}} \frac{z^n}{n} a_n\:,\:\:\:\:\:
\hat{\phi}(\overline{z}) = \sum_{n\in \bZ\setminus \{0\}} \frac{\overline{z}^n}{n} b_n$$
where the operators $b_n$ are defined on $\sP$ similarly to  operators $a_n$ and
$[a_n,b_m] = 0$. The operators $F_n$ and the
analogues $P_n$ defined in ${\mathsf{P}}$ are those usually denoted by $L_n$ and $\overline{L_n}$ respectively. \\
{\bf (3)} The bulk evolution is generated by the Hamiltonian $H$  which is the quantum generator associated with
the bulk killing vector $\partial_t$.  Consider  the operator $H_{\sF}$ associated with $H$
by holography  and 
naturally extended it in the whole Fock space ${\gF}({\cal H}_\sF)$ by assuming to work with massive non interacting
particles in the bulk. The obtained operator $H^\otimes_\sF$
coincides with the self-adjoint Virasoro generator
\beq H^\otimes_\sF :=\frac{1}{\beta}\overline{\left(2F^{(+)}_1 + F_0\right)}\:\:\:\:\:\:\:\: \:\:\:\:\:\:\:\:\:\:\:\:\:\:\:\: \:\:\:\:\:\:\:\:\mbox{(where $\mu=0$)}\:.
\label{Htensor}\eeq
{\em provided  $\mu=0$}.  Under this hypothesis,  $F^{(+)}_0, F^{(\pm)}_1$ span a finite-dimensional 
Lie algebra and $H^\otimes$ is the closure of an element of the algebra. (That is noting but the Lie algebra of 
a unitary representation of $SL(2,\bR)$.)
As a consequence it is possible to define time-depending observables $F_0(t), F^{(\pm)}_1(t)$ which are constant 
of motion in Heisenberg picture. These are  finite linear combinations of generators $F_0(t), F^{(\pm)}_1(t)$. 
The proof of that fact  is essentially the same as that of  Theorem 2.1 -- item  (b) in particular -- in \cite{mopi03}. 
We conclude that  $F_0(t), F^{(\pm)}_1(t)$ generate a {\em symmetry of the system} when they are realized, 
by   unitary holography, as operators acting in the bulk.
 Conversely, this result does not apply as it stands  for $F^{(\pm)}_n$ if $n>1$. This is because there is no finite-dimensional Lie algebra
 containing both $F^{(\pm)}_n$ and $H^{\otimes}_\sF$. However
 if one assumes that $F_0^{(+)}$ (which is associated with $K_{\beta \bF}$ in the bulk) 
  is the Hamiltonian of the theory on $\sF$, the observables $F_n^{(\pm)}$ with $n>1$ can be considered 
  as symmetries of the system. This is  because, for every fixed integer $n>0$ and also if $\mu\neq 0$,
   $F_0^{(+)}, F_n^{(+)}, F_n^{(-)}$ span a finite-dimensional Lie algebra
  (which is, in fact, a representation of the Lie algebra of $SL(2,\bR)$).

\section{Appearance of thermal states from Virasoro generators.}

 \noindent Let us focus on the class of  ``Hamiltonian operators'' defined for the theory on the circle
 $\sF$,
 \begin{eqnarray}
 H^\otimes_{\sF\:\mu} :=\frac{1}{\beta}\overline{\left(2F^{(+)}_1 + F_0\right)}\:,
 \end{eqnarray}
 where, differently from (\ref{Htensor}),  now $\mu\in \bR$ and thus, barring the value $\mu=0$, $H^\otimes_{\sF\:\mu}$ 
 cannot be associated with the Rindler Hamiltonian in the bulk by means of  holography.
 In the following we study some properties of these Hamiltonians and associated ground states
 which can  be considered as operators and states  of the theory in the bulk.
 We shall not give rigorous proofs since the treatment  of the issue involves a
 singular Bogoliubov transformation as well as a regularization procedure. 
 We have formally
\beq
H^\otimes_{\sF\:\mu} =\frac{\mu^2}{2\beta}I+ H^\otimes_\sF +i\frac{\mu}{2\beta}\at\alpha_1-\alpha_1^\dagger\ct.
\eeq
 We look for a, formally unitary, transformation $U_\mu$ 
 such that
 $$ H^\otimes_\sF  =U_\mu H^\otimes_{\sF\:\mu}U_\mu^\dagger \:.$$
 It is convenient to work in the Fock space $\gF({\cal H}_\bF)$ which is isomorphic to $\gF({\cal H}_\sF)$
 by means of the isomorphism $M : {\cal H}_\bF \to {\cal H}_\sF$ used in Theorem 2.1.
 In this representation\beq
H^\otimes_{\sF\:\mu} = \frac{\mu^2}{2\beta}I + \int_{\bR^+} dE\: a^\dagger_E a_E
+i\frac{\mu}{2\beta}\int_{\bR^+} dE\: Z^{(1)}_1(E)\at a_E-a_E^\dagger\ct\:,
\eeq 
where $a_E$ and $a_E^\dagger$ are as in eq. (\ref{nonrig}).
 By that way, it turns out that formally
 \beq
U_\mu=\exp{\at-i \int_0^\infty Z^{(1)}_1(E)\frac{\mu}{2\beta} (a_E+a_E^\dagger)\frac{dE}{E} \ct }\:. \label{start}
\eeq
Notice that when $\mu$ is equal to zero the unitary transformation becomes the identity and $H^\otimes_{\sF\:0}=H^\otimes_\sF$ as is due.
For completeness we say that it is possible to rewrite $U_\mu$  in terms the operators $\alpha_n$ as
follow:
\beq
U_\mu=\exp{\at-i \sum_{n>0}(-1)^{(n+1)}\frac{\mu}{\sqrt{n}}(\alpha_n+\alpha_n^\dagger)\ct }\:.
\eeq
The ground state of $H^\otimes_{\sF\:\mu}$, can be obtained as
\beq
\Psi_\mu := U^\dagger_\mu |0\rangle_{\sF}\:.
\eeq
$\Psi_\mu$ is not invariant under Rindler evolution generated by $H^\otimes_{\sF}$ but it enjoys interesting thermal  properties when one considers
expectation values of observables also averaged during  a long period of Rindler time $T\to \infty$.
 Consider the expectation value  of the operator $A$:
\beq
\langle A\rangle_\mu : =\lim_{T\to\infty}\frac{1}{2T} \int^T_{-T}\frac{\langle \Psi_\mu(t), A \Psi_\mu(t)\rangle_\sF}{\langle \Psi_\mu(t),
 \Psi_\mu(t)\rangle_\sF} dt \label{AM}\:,
\eeq
 where $\Psi_\mu(t) := \exp\{-it H^{\otimes}_{\sF}\}\Psi_\mu$. The direct computation is affected by 
mathematical problems  which can be made harmless by making discrete the energy spectrum 
 and  taking the limit toward the continuous case after the evaluation of the expectation value.
 The discrete spectrum can be obtained by reducing to a known regularization procedure,
 consisting into a suitable version of  the so-called  ``box quantization''. Start from (\ref{start}) noticing that it can be re-written, by using
 the adimensional parameter $\lambda= \log(E/E^*)$,
$E^*$ being an arbitrarily fixed  energy scale,
 \beq
U_\mu=\exp{\at-i \int_{-\infty}^{+\infty} {Z^{(1)}_1(E(\lambda))}\frac{\mu}{2\beta} \at a_{E(\lambda)}+a_{E(\lambda)}^\dagger\ct d\lambda \ct }\:.
\label{start'}
\eeq Now, differently from $E$, $\lambda$ ranges over the whole real line and box-quantization can be used as follow.
First of all   define the operators $c_\lambda=\sqrt{E'(\lambda)} a_{E(\lambda)}$
and  $c^\dagger_\lambda=\sqrt{E'(\lambda)} a^\dagger_{E(\lambda)}$ 
so that bosonic commutation relations of $a_E$ and $a^\dagger_E$
turns out to be  equivalent to 
\beq
\aq  c_\lambda, c_{\lambda'}^\dagger \cq=\delta(\lambda-\lambda')\:,\:\:\:\:\:  \aq c_\lambda, c_{\lambda'} \cq=0
\:,\:\:\:\:\:  \aq c_\lambda^\dagger, c_{\lambda'}^\dagger \cq=0\:.\eeq
Finally, to get the discrete  spectrum in $\lambda$, assume that values  $\lambda$ describe  the spectrum of a ``momentum operator''.
These values  can be made discrete by working  a  $1D$ box with length $L$
with periodic boundary conditions, the continuous spectrum being restored in the limit $L\to \infty$.
Within this framework, if $\lambda_n = \frac{2\pi n}{L}$ with $n\in\mathbb{Z} $,
the operators $c_j := c_{\lambda_j}$ enjoy the commutation relations
\beq
\aq  c_i, c_{j}^\dagger \cq=\delta_{ij}\:,\:\:\:\:\:  \aq c_i, c_{j} \cq=0
\:,\:\:\:\:\:  \aq c_{i}^\dagger, c_{j}^\dagger \cq=0\:.\eeq
With that regularization procedure, the Hamiltonian $H^\otimes_{\sF}$ can be re- written as
\beq
H^\otimes_{\sF}=\int_{\bR^+}  E\; a_E^\dagger\; a_E\; dE=\int_{\bR} E(\lambda)\; c_\lambda^\dagger\; c_\lambda
\: d\lambda \to \sum_j E_j c^\dagger_jc_j,
\eeq
where $E_j:=E(\lambda_j)$.
Similarly, using $E=E^*e^\lambda$ and  (\ref{ZEgen}) for $k=n=1$, the regularized unitary transformation $U_\mu$ reads:
\beq
U_\mu = \prod_j \exp \at-i \mu e^{-\beta E_j}(c_j+c_j^\dagger)\ct.
\eeq
and so  the state $\Psi_\mu$ can be expanded as 
\beq
\Psi_\mu=\prod_j \exp{\at \frac{\mu^2}{2} e^{-2\beta E_j}\ct}\;\sum_n i^n\mu^n e^{-\beta E_j n}
\frac{{c^\dagger_j}^n}{n!} |0\rangle_\sF. \label{UP}
\eeq
We are now ready to compute $\langle A\rangle_\mu$. Using (\ref{UP}) in (\ref{AM})
one gets straightforwardly 
\beq
\langle A\rangle_\mu = Z_{\beta}^{-1}\sum_{\ag n_j\cg} e^{-2\beta\sum_j E_j
n_j} \mu^{2\sum_jn_j} \left\langle \{n_j\}|A|\{n_j\}\right\rangle, \label{Fine}
\eeq
with
\beq
 Z_{\beta} = \sum_{\ag n_j\cg} e^{-2\beta\sum_j E_j
n_j} \mu^{2\sum_jn_j}\:,
\eeq 
and the final limit $L\to \infty$ is understood.
Let us consider all the developed machinery as referred to the theory in the bulk
making use of the holographic theorem (Theorem 2.2) for a {\em massive} field.
In that way, $A$ must be considered  as an observable for an observer in the Rindler wedge and $\Psi_\mu$
is a state for a quantum field propagating in the Rindler wedge. 
If $\mu$ is equal to $1$, (\ref{Fine}) states that the time-averaged state $\Psi_{\mu=1}$ 
viewed by an observer in the bulk who uses Rindler time-evolution, is a thermal state
with inverse temperature $1/(2\beta)$. In particular,
if we also  choose $\beta=\beta_U/2$, where $\beta_U$ is the inverse Unruh temperature, 
we get formally and in the sense of the pointed regularization-procedure out,
\beq\label{average}
\langle A\rangle_{\mu=1} = tr \at \rho_{\beta_U} A\ct,
\eeq
where $$\rho_{\beta_U} :=  \frac{e^{-\beta_U H^\otimes_\sF}}{ tr e^{-\beta_U H^\otimes_\sF} } $$ is the  density matrix of a thermal state,
which coincides with the restriction of  Minkowski vacuum
to Rindler wedge because of celebrated results of QFT (Bisognano-Wichmann-Sewell theorems, see \cite{haag}
for a general discussion). In the case $\mu\neq 1$, 
(\ref{Fine}) suggests to interpret  $$\frac{\log \mu^2}{2 \beta}$$ as a chemical potential and the associated state can be seen as a grand canonical ensemble state.

\section{Overview and open problems} 

In this paper we have shown that quantum field theory for free fields propagating in  a $2D$-Rindler background is  unitary  equivalent
to the analogue defined on the compactified Killing horizons.
The same equivalence can be implemented at algebraic level. 
 The key point of this holographic description 
is the  hidden $SL(2,\bR)$ symmetry found in \cite{mopi02} for the fields propagating in the bulk. Indeed that hidden representation of $SL(2,\bR)$
 becomes geometrically manifest when
the theory is represented on the Killing horizon.  Preserving a clear geometric meaning, the representation  can be enlarged up to include a
positive-energy
unitary representation of Virasoro algebra with central charge $c=1$.
Notice that the Virasoro algebra is realized in the many particles description of the fields, namely it
describes a representation in the Fock space. The appearance of the pair of Virasoro algebras in the future and past horizon
leads naturally to an (Euclidean) $2D$ conformal field theory on a cylinder which is holograpically associated with QFT in the bulk.
In the last section we have proposed  the idea that, for a particular choice of the parameter $\beta$ and the ground energy $h=\mu^2/2$
of the Virasoro Hamiltonian $F_0$,  the ground state $\Psi_\mu(t)$ of another Virasoro generator which generalizes Rindler Hamiltonian
has thermal properties. $\Psi_\mu(t)$, seen in Fock space built up over the Rindler vacuum $|0\rangle$, reveals to be an infinite particle
state in thermal equilibrium temperature $1/(2\beta)$.
It can be useful to describe the Hawking effect.
These thermal properties are  shown here without rigorous proof 
because of the use of  a necessary regularization procedure  in computing the mean
value of the state with respect to 
Rindler time.  Further investigation in that direction, perhaps based on KMS condition, is necessary.\\
Another issue which deserves investigation is the existence of any relation between the results of this paper
and the attempts to give a statistical explanation to black hole entropy by counting microstates of irreducible unitary representations of
Virasoro algebra \cite{carlip-etall}. This is done  by means of the so-called Cardy's formula after a suitable dimensional 
regularization which gives rise to a scalar field (supporting part of information 
of $4D$ gravity)  propagating in a $2D$ spacetime. The main problem of those approaches is that they must assume the existence
of a quantum Virasoro representation. The existence of such a representation has been established in this paper: It is worthwhile to investigate about the possible interplay
between the quantum Virasoro representation found here and that necessary in those approaches.

\section*{Acknowledgments}

N.P. would like to thank the Department of Physics of the University of Trento (especially Guido Cognola
and Sergio Zerbini) for  kind hospitality during the completion of this work.

\begin{center}
\epsfig{file=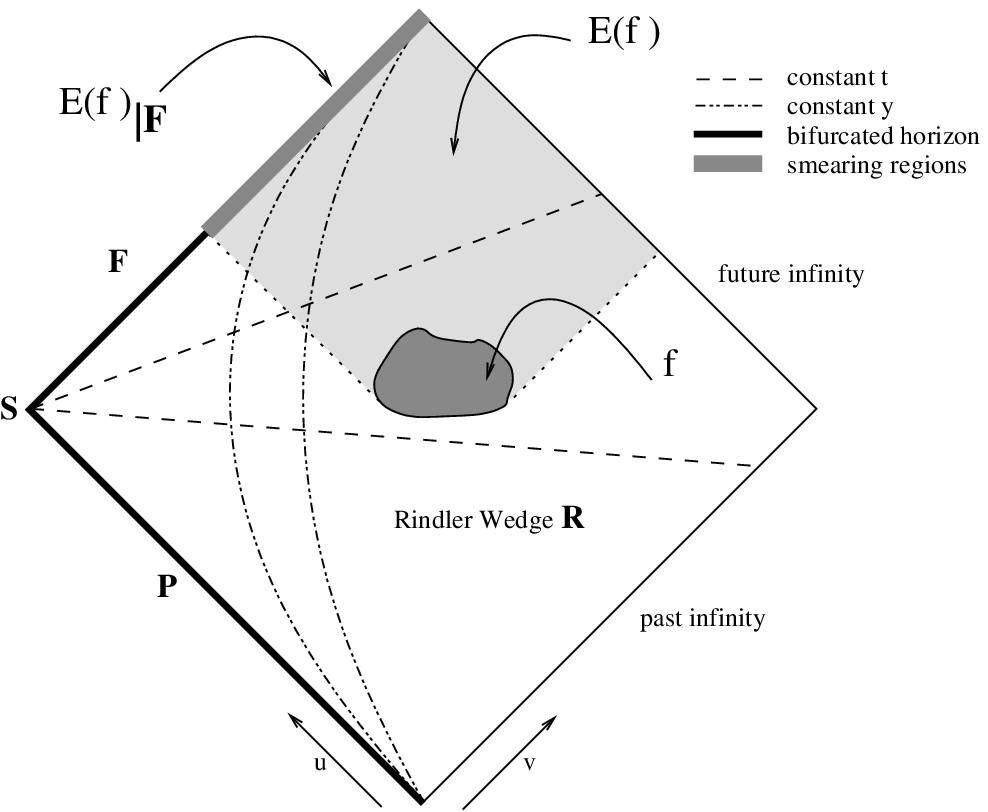,width=1\textwidth} Holography in $2D$-Rindler space.
 \end{center}

\end{document}